\newif\ifblind
\blindfalse

\ifblind
\documentclass[conference,compsoc]{IEEEtran}
\else
\documentclass[sigconf, nonacm]{acmart}
\fi

\AtBeginDocument{%
  }

\usepackage{hyperref}
\usepackage{booktabs}
\usepackage{graphicx}
\usepackage{float}
\usepackage{amsmath}
\usepackage{xspace}

\usepackage{amssymb}
\usepackage{subcaption}
\usepackage{enumitem}
\usepackage{quoting}
\usepackage{algorithm}
\usepackage{titlesec}
\usepackage[noend]{algpseudocode}
\usepackage{caption}
\usepackage{stfloats}
\usepackage{placeins}
\usepackage{xurl}

\graphicspath{{figs/}}
\DeclareGraphicsExtensions{.png,.pdf}

\newcommand\mytt[1]{\texttt{\small{#1}}}
\newcommand\xdo{$D_{orig}$\xspace}
\newcommand\xdop{$D'_{orig}$\xspace}
\newcommand\xda{$D_{anon}$\xspace}
\newcommand\xdap{$D'_{anon}$\xspace}
\newcommand\xpa{$P_{atk}$\xspace}

\newcommand\xprca{$PRC_{atk}$\xspace}
\newcommand\xprcab{$PRC_{atk}^{best}$\xspace}
\newcommand\xprcb{$PRC_{base}$\xspace}
\newcommand\xprcbb{$PRC_{base}^{best}$\xspace}
\newcommand\xcm{$R_{min}$\xspace}

\newcommand\rs{\textit{rank\_score}\xspace}
\newcommand\tup{\textlangle\textit{prediction,rank\_score\textrangle}\xspace}
\newcommand\atin{attribute inference\xspace}

\ifblind
\newcommand\codeurl{https://github.com/anon-paper-submissions1/alc-paper}
\newcommand\datasetsurl{https://github.com/anon-paper-submissions1/alc-paper}
\else
\newcommand\codeurl{https://github.com/yoid2000/arxiv-alc-paper-2025}
\newcommand\datasetsurl{https://github.com/yoid2000/arxiv-alc-paper-2025}
\fi

\begin{document}

\title{Towards Better Attribute Inference Vulnerability Measures}

\ifblind
\else

\author{Paul Francis}
\orcid{0009-0000-1574-887X}
\affiliation{%
  \institution{MPI-SWS}
  \city{Kaiserslautern}
  \country{Germany}}
\email{francis@mpi-sws.org}

\author{David Wagner}
\orcid{0009-0000-2786-8158}
\affiliation{%
  \institution{Deutsche Universit{\"a}t f{\"u}r Verwaltungswissenschaften}
  \city{Speyer}
  \country{Germany}}
\email{dwagner@uni-speyer.de}

\fi

\ifblind
\else
\renewcommand{\shortauthors}{Francis et al.}
\fi


\begin{abstract}
The purpose of anonymizing structured data is to protect the privacy of individuals in the data while retaining the statistical properties of the data. An important class of attack on anonymized data is attribute inference, where an attacker infers the value of an unknown attribute of a target individual given knowledge of one or more known attributes. A major limitation of recent attribute inference measures is that they do not take recall into account, only precision. It is often the case that attacks target only a fraction of individuals, for instance data outliers. Incorporating recall, however, substantially complicates the measure, because one must determine how to combine recall and precision in a composite measure for both the attack and baseline. This paper presents the design and implementation of an attribute inference measure that incorporates both precision and recall. Our design also improves on how the baseline attribute inference is computed.  In experiments using a generic best row match attack on moderately-anonymized microdata, we show that in over 25\% of the attacks, our approach correctly labeled the attack to be at risk while the prior approach incorrectly labeled the attack to be safe.

\end{abstract}
\maketitle

\section{Introduction}
\label{sec:introduction}

An important and heavily-researched problem is that of anonymizing structured data. The goal is to preserve the statistical properties of the data while protecting the anonymity of individual persons in the data. A critical aspect of the study of data anonymization is the ability to \textit{measure} anonymity. Over-estimates of anonymity could lead to the release of unsafe data, while under-estimates could lead to overly aggressive anonymization, resulting in unnecessarily poor data utility.

While there are many measures of anonymity~\cite{wagner2018technical}, an important measure is attribute inference: the ability of an attacker to predict the value of an unknown attribute for a target individual. Attribute inference is one of the three criteria for evaluating anonymity in the Article 29 Data Protection Working Party Opinion 05/2014 on Anonymisation Techniques~\cite{article29}, though the Opinion does not give any guidance on how to recognize when excessive inference has occurred. The use of \atin as a measure has increased in recent years with concerns over the privacy of ML models~\cite{fredrikson2014privacy,yeom2018privacy} and synthetic data~\cite{stadler2020synthetic,giomi2022unified}.

A central problem with \atin is that of distinguishing when \atin is a true privacy loss and when it is simply a legitimate statistical inference. For example, given an anonymized demographic dataset of professions, suppose there is an attack predicting sex that achieves 98\% accuracy. This may seem like an effective privacy-violating attack, but if the class of individuals being attacked is ``construction worker'', the ``attack'' is to simply predict male, and 98\% of construction workers are indeed male, then the attack is in fact privacy-neutral.

\begin{figure}
\begin{center}
\includegraphics[width=1.0\columnwidth]{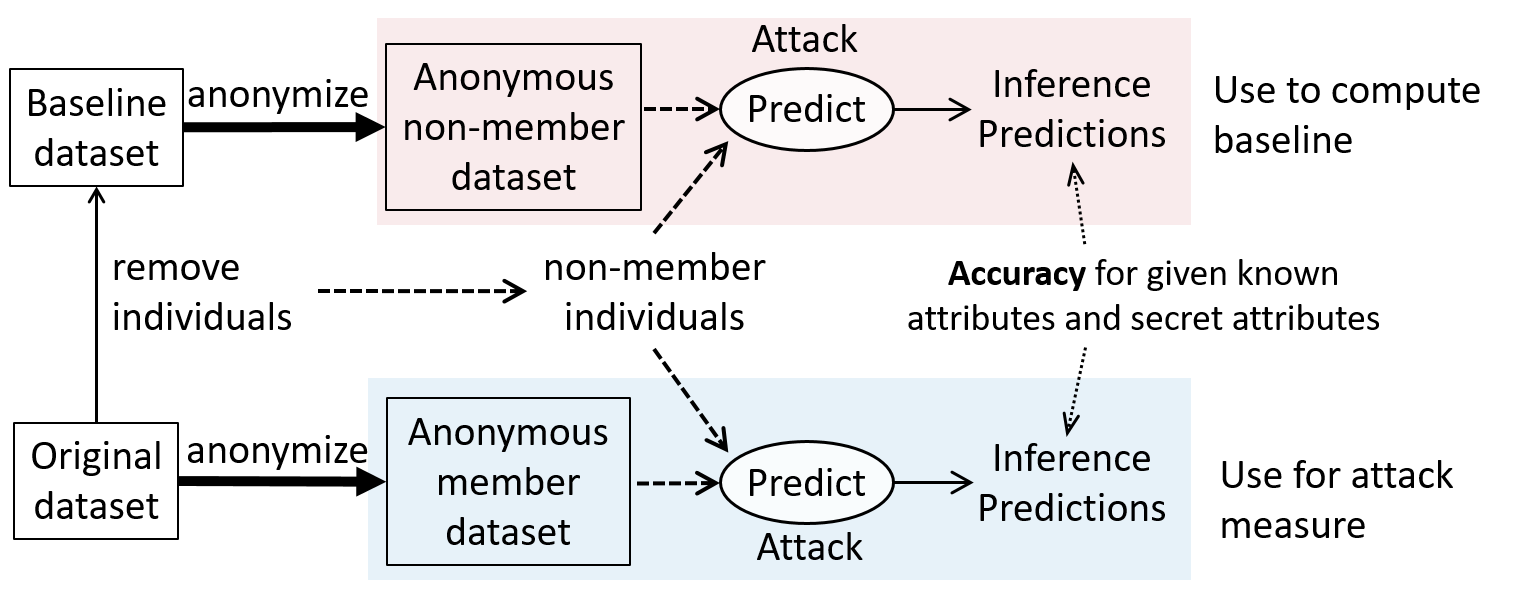}
\caption{Prior approach to computing an \atin baseline. Note that both the member and non-member datasets are anonymized, and that the same attack is used on both datasets to predict attributes. Prior approaches compute accuracy, and do not take recall into consideration.}
\label{fig:prior-approach}
\end{center}
\end{figure}

It is therefore necessary to establish an \atin privacy-neutral \textit{baseline}. This is the quality of \atin that applies to the general population, and not to specific individuals. An elegant way to establish this baseline has emerged in recent years~\cite{yeom2018privacy,kassem2019differential, stadler2020synthetic, giomi2022unified, kifer2022bayesian}. The idea is to produce two anonymized datasets that differ by the presence of a single individual (the \textit{target}). We refer to these as the \textit{member} and \textit{non-member} datasets. An \atin attack on the target is run against both anonymized datasets. This procedure is repeated over enough targets to produce high-confidence accuracy measures for both member and non-member datasets. The improvement in accuracy for the member dataset over the non-member dataset represents the privacy loss due to releasing the anonymized dataset for the given attack. This basic approach is illustrated in Figure~\ref{fig:prior-approach}.

The underlying principle behind this approach is that a dataset cannot be said to violate the privacy of individuals who are not part of the dataset and who are independent from other members of the dataset. This is a core principle behind differential privacy, initially stated in the original paper~\cite{Dwork06} (see discussion of Terry Gross' height), and reiterated in~\cite{dwork2017exposed}.


\textit{Higher-accuracy non-member inferences constitute better baselines.} To see why, consider a professions dataset where the following two techniques to predict the sex of a non-member known to be a construction worker. The first technique bases the prediction only on the distribution of the sex attribute. Say this leads to a 50\% accuracy baseline. The second technique is to take advantage of the fact that most construction workers are male, and to always predict male. Say this leads to a 98\% accuracy baseline. Now suppose that the professions dataset is anonymized and released to the public, and that an "attacker" again always predicts male for construction workers that are in the anonymized dataset. If the lower 50\% baseline is used, then this attack would wrongly be perceived as compromising privacy because the attack accuracy is much higher than the baseline accuracy. If the higher 98\% baseline is used, then the attack would rightly be regarded as not compromising privacy. In the former case, useful data that should otherwise be released to the public might be withheld.

This paper presents a new design for \atin measures that improves on prior work in two ways. First, our design measures \textit{precision} and \textit{recall}, whereas prior work only measures accuracy, and ignores recall. Attacks that have low accuracy across all individuals, but high accuracy for a small fraction of individuals are common.  For example, Narayanan and Shmatikov reported that reidentification confidence for only 2 of 50 targeted individuals in the Netflix dataset was very high~\cite{narayanan2008robust}. Although they don't provide numbers, we can suppose that an accuracy measure taken across all 50 individuals might well be quite low and therefore regarded as safe, whereas an accuracy measure taken over 2/50=4\% of attacked individuals is very high and therefore unsafe for that 4\%. Similarly, Cohen reported only 3 high-confidence reidentifications of 135 targets in the edX online classes dataset~\cite{cohen2022attacks}, and Abowd et al. reported that less than 1\% of reconstructions led to high-confidence predictions~\cite{abowd2023census}.

Although none of these three studies use the term \textit{recall} per se, they all convey that higher accuracy can be achieved over a subset of the population, and that this subset cannot be identified prior to the attack, but rather from information derived during the attack itself. This suggests an approach whereby a confidence score is associated with each prediction, and a family of precision/recall measures are computed by applying different thresholds to the confidence scores.

Recall complicates the use of baselines because both the baseline and attack measures have both precision and recall components. Further, the recall values may be different between attack and baseline, complicating any direct comparison of the two precision components. To deal with this, we define a composite measure, the Precision Recall Coefficient (PRC), that allows a direct comparison between the baseline and the attack measures, and the resulting measure of anonymity loss.

The second improvement is that our approach generates more accurate baselines, and therefore better measures. This is accomplished with two modifications to the prior approach. First, we don't anonymize the non-member dataset. Since anonymization usually reduces data quality, and therefore reduces the quality of non-member inferences, this modification improves the baseline. Second, we don't use the attack itself to compute the baseline, but rather we allow the use of any predictive analytic technique (including the attack itself, if that happens to produce the best baseline).

\begin{figure}
\begin{center}
\includegraphics[width=1.0\columnwidth]{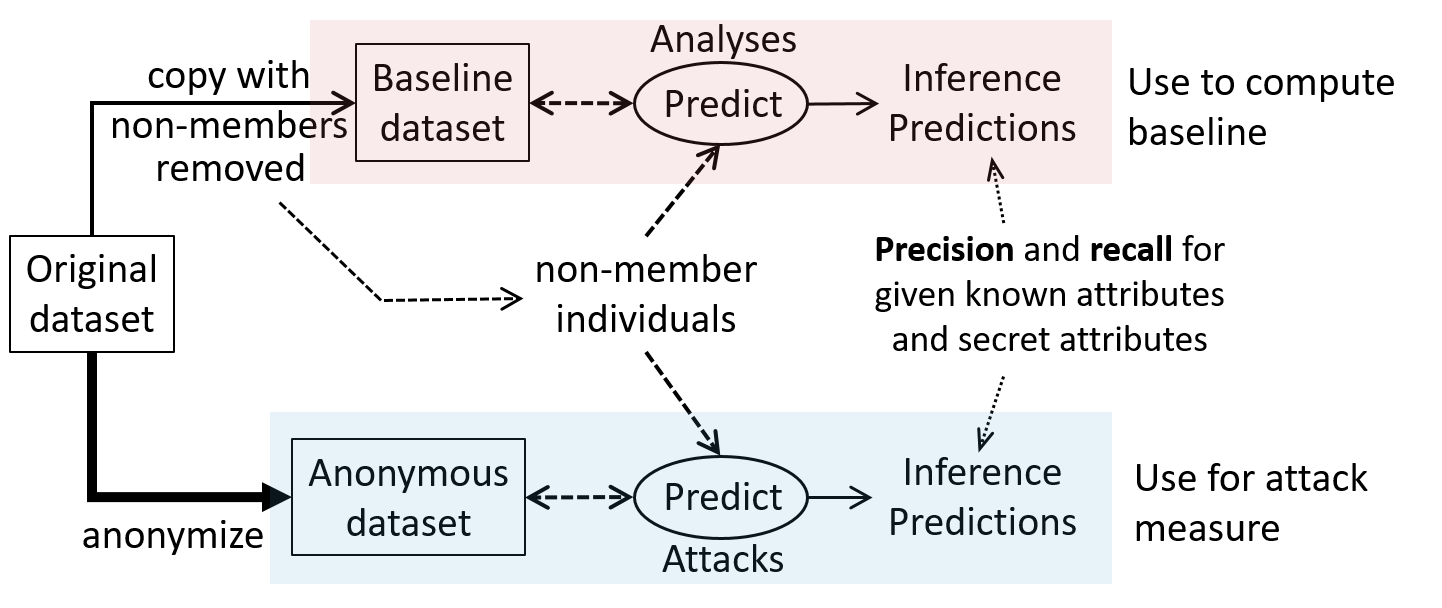}
\caption{Our approach to computing an \atin baseline. Compared to the prior approach of Figure~\ref{fig:prior-approach}, our approach does not anonymize the non-member dataset, and allows the use of any predictive analytic technique to compute the baseline. Our approach computes precision and recall.}
\label{fig:non-member}
\end{center}
\end{figure}

As shown in Section~\ref{sec:prior-approaches}, these improvements, when applied to an attack on a dataset with moderate anonymization, label over 25\% of the attacks as being at risk which the prior approach labeled as being safe. Our approach is illustrated in Figure~\ref{fig:non-member}.



Another advantage of using the original data for the baseline is that it is more efficient, since it skips the anonymization step for the baseline. For instance, the anonymization step was considered to be prohibitively expensive in the US Census Bureau's second analysis of the reconstruction attack~\cite{abowd2023census}, and so they were forced to use an approximation of the baseline instead.

As a point of clarification, we stress that the intent of the baseline is not to model what a real attacker may or may not know in practice (an infeasible task in any event), but rather to establish an \textit{ideal, theoretical} baseline: one that would be privacy-neutral even if the attacker had all of the original data minus the target individual.

The contributions of this paper are:
\begin{itemize}
\item A new design for measuring the vulnerability of \atin attacks that measures recall as well as precision, and is more accurate and more efficient than prior approaches.
\item An implementation of our measure as a pip-installable python package (\href{https://github.com/yoid2000/anonymity_loss_coefficient}{GitHub repo \mytt{anonymity\_loss\_coefficient}}), including a framework for adding attacks.
\item The design and implementation of a general attack for use with synthetic data that improves on Giomi et al.~\cite{giomi2022unified} attack by taking into account the quality of the best matches.
\item Experiments showing the improvement in performance or our design over prior approaches. In so doing, we also show that it is important to include weakly-anonymized datasets in the evaluation of \atin attacks.
\end{itemize}

Section~\ref{sec:related} describes related work, and in particular discusses flaws in prior vulnerability measures, including that of the US Census Bureau's reconstruction attack. Section~\ref{sec:design} describes our design for the baseline and attack measures. Section~\ref{sec:composite} describes how the baseline and attack measures can be combined into composite measures that allow direct comparison between measures that have different recall. Section~\ref{sec:implementation} describes the implementation of our approach, including the design of a best row match attack that exploits recall.  Section~\ref{sec:prior-approaches} describes our experiments and results, comparing our approach with the prior approach. Section~\ref{sec:summary} summarizes the paper and outlines future work.
\section{Related work}
\label{sec:related}

There are dozens of anonymity measures. Good surveys exist for statistical disclosure~\cite{matthews2011}, membership~\cite{hu2022membership}, location~\cite{liu2018location}, synthetic data~\cite{figueira2022survey}, and differential privacy~\cite{desfontaines2020sok}.

Much of the early work on data anonymity in the mid 70s to early 80s used attribute inference as a vulnerability measure~\cite{fellegi1974statistical, dalenius1977towards, denning1979tracker, denning1980secure}. Much of this work focused on intersection and tracking attacks on noiseless anonymization techniques (aggregation and suppression), and so successful attacks would have 100\% precision. Such attacks are a clear vulnerability regardless of the baseline, and so no baseline was needed in this early work.

A prominent attack in the literature is the Dinur-Nissim reconstruction attack in 2003~\cite{dinur2003revealing}. Although the vulnerability measure for this attack is the percentage of correct reconstructions, a correct reconstruction can also be seen as a correct attribute inference prediction, where the attribute takes on the values 0 or 1. Since the values are uniformly distributed, the baseline precision is understood to be 50\%, and so no special technique to compute baseline was needed.

In 2014, two independent works recognized the need for a baseline when measuring \atin. Elliot studied anonymization of survey data using the Synthpop package to produce synthetic data~\cite{elliot2015final}, and Fredrikson et al. studied the release of an ML model used to predict the dosage of the drug warfarin~\cite{fredrikson2014privacy}. In both cases, the authors used the univariate distribution of the unknown attribute as the baseline. This does not produce the best baseline, since it does not take advantage of information from known attributes. Therefore, both studies overestimate vulnerability.

In the case of Elliot, this didn't really matter as Elliot had in any event found the synthetic data to be safe. By contrast, Fredrikson et al. had concluded that the ML model was vulnerable to a model inversion attack. Ironically, Fredrikson et al. had also run the inversion attack on a holdout (validation) set, without recognizing that this would serve as a better baseline. This reduced the precision improvement over the baseline from 22\% using the univariate distribution as the baseline, to only 3\% using the holdout set as the baseline. As a result, we can conclude that Fredrikson et al. had drawn the wrong conclusion about the privacy risk of the ML model.

In 2018, Yeom et al. proposed a non-member approach to evaluating privacy leakage from model inversion attacks on released ML models~\cite{yeom2018privacy}. They compare the attribute prediction precision of running the model inversion attack against members of the model's training set with the attack run against members of the test set (non-members). This is to our knowledge the first use of a non-member baseline for \atin.

Kassem et al. (2019~\cite{kassem2019differential}) propose a general framework for evaluating anonymity whereby the anonymization of two datasets, one with and one without a target individual, are compared using any distance function. This distance function could in principle be based on attribute prediction, though Kassem et al. demonstrate a different distance function.

Stadler et al. (2020~\cite{stadler2020synthetic}) and Giomi at al. (2022~\cite{giomi2022unified}) both evaluate the effectiveness of their attacks on synthetic data by comparing predictions made on members of the synthetic dataset with predictions made on non-members. Kifer et al. (2022~\cite{kifer2022bayesian}) suggests a non-member approach to evaluating the privacy of US Census data in the context of differential privacy.

Though details differ, all of the above cited works effectively use an approach similar to that of Figure~\ref{fig:non-member}. None of them take recall into consideration.

In 2019, Abowd published the results of a reconstruction attack on portions of the US Census Bureau (USCB) 2010 Decennial Census data~\cite{abowd-2019-staring}. The attack takes tabular (aggregate) data published by the USCB, attempts to reconstruct the original records, and then reidentifies individuals in the reconstructed records by linking them with public data. Abowd stated that 46\% of the population could be correctly reconstructed, and that 17\% of the population could be correctly reidentified from this attack. In 2021, Ruggles and Van Riper showed, however, that similar reconstruction quality could be achieved simply by the random assignment of attributes taken from national-level statistics~\cite{ruggles2021role}. The Ruggles and Van Riper random reconstruction is itself a kind of baseline, albeit one based on reconstruction rather than \atin. This strongly suggested that the Abowd reconstruction attack grossly overestimated vulnerability by failing to take into account a baseline reconstruction.

Altman et al. also regard random reconstruction (or random guessing) as a baseline~\cite{altman2021hybrid}.

In 2021, Abowd published additional details on the reconstruction attack in response to a lawsuit filed against the USCB by the state of Alabama~\cite{lawsuit-alabama}. In addition to reconstruction and reidentification measures, Abowd also reported \atin measures, where race/ethnicity are the unknown attributes, and age, sex, and census block (geographic area) are the known attributes. Abowd argued that the precision and recall of the \atin measures demonstrated that the 2010 Decennial Census data was highly vulnerable. In 2022, however, Francis demonstrated that even better levels of precision and recall could be achieved by simply predicting that every member of a block has the majority race/ethnicity of the block~\cite{francis2022census}. This simple assignment of the majority race/ethnicity represented a kind of \atin baseline, since only statistics about the general population are being used to make the inferences. According to this baseline, the Abowd reconstruction attack demonstrated no vulnerability whatsoever.

In 2023, Abowd et al. revised its analysis of the reconstruction attack to address the errors in the earlier reports~\cite{abowd2023census}. The revised analysis sketches out a non-member approach to establishing a baseline, again along the lines of prior work as illustrated in Figure~\ref{fig:non-member}. The revised analysis does not implement that baseline, however, on the grounds that it would be too expensive to compute. Instead, they first simply define individuals with the majority race/ethnicity in their respective blocks as no longer vulnerable (called \textit{modals}). Then they focus their analysis on the remaining individuals (\textit{non-modals}), and define a baseline as a random guess of race/ethnicity from the race/ethnicity distribution among non-modals in each given block. Using this approach, they find that they can get 95\% precision predicting race/ethnicity on 0.3\% of the population.


In addition to the prior work cited above, note that the idea of comparing datasets that differ by a single individual can be used to measure differential privacy. There are many examples of this, but Gehrke et al. in particular is similar to our non-member framework in that it generates a baseline dataset from the (sampled) original data rather than from anonymized data~\cite{gehrke2011towards}. This work is used to provide a definition of differential privacy rather than generate a baseline per se. Other examples can be found in~\cite{desfontaines2020sok}. More recently, empirical evaluation of DP mechanisms has been proposed to generate tighter privacy bounds under less pessimistic attacker assumptions~\cite{jagielski2020auditing,nasr2021adversary,steinke2023privacy}.

\section{Design of our anonymization measure}
\label{sec:design}

The setting is that a data custodian wishes to release information about a dataset \xdo  in anonymized form. The release could be a single table, as with synthetic data or the release of a K-anonymized dataset, or could be multiple tables or statistics as is commonly done by statistics agencies, or could even be a dynamic query interface to an anonymizing system. The custodian wishes to measure the strength of anonymity of the released data \xda.

The data custodian has one or more \textit{attack scenarios}. Each scenario is defined by:
\begin{itemize}[leftmargin=0.5cm]
 \item Zero or more known attributes $KA_t$ for each target individual $t$
 \item Unknown attributes $UA_t$ for each target individual $t$
 \item A target selection function \\ \hspace*{0.3cm} $select_{atk}$(\xdo, \textit{mode}) $\rightarrow$ \textit{targets} \\ \hspace*{0.3cm} where \textit{mode} is \textit{pre-targeted} or \textit{random}
 \item A predictor for the attack \\ \hspace*{0.3cm} $pred_{atk}$(\xda, $KA_t$, $UA_t$) $\rightarrow$ [\textit{T, F}], \rs \\ \hspace*{0.3cm} where \xda is the anonymized dataset
 \item A predictor for the baseline \\ \hspace*{0.3cm} $pred_{base}$(\xdop, $KA_t$, $UA_t$) $\rightarrow$ [\textit{T, F}], \rs \\ \hspace*{0.3cm} where \xdop is the original dataset minus the target 
\end{itemize}

\subsection{Predictors}
\label{sec:predictors}

\textit{Predictors} ($pred_{atk}\text{()}$ or $pred_{base}\text{()}$) return two values, the true/false result of the prediction, and an optional ranking score \rs. The predictor uses the supplied dataset (\xda or \xdop) and the known attributes $KA_t$ to predict the unknown attribute. If the predicted unknown attribute matches that of the target $UA_t$, then the predictor returns true, otherwise false.

We use the term \textit{attribute} broadly. Normally we might think of it as a column in a dataset that takes on a value or range of values for a given individual. But known attributes $KA_t$ could be any set of facts that are true for target individual $t$ (has visited the hospital 10 times, is between age 20 and 25, and has never been to Paris). Likewise unknown attributes $UA_t$ can be any logical expression that evaluates to true or false (had cancer, lives in zip code 12345, 12346, or 12347, and frequently visits London).

The ranking score \rs is a value that reflects the predictor's estimated quality of the prediction relative to other \rs values from the same predictor. A higher \rs is estimated to be more likely correct than a lower \rs. The \rs is only used to rank predictions from a given predictor. The value itself is not necessarily a probability of correctness, and a \rs of $pred_{atk}\text{()}$ cannot be compared with a \rs of $pred_{base}\text{()}$. The \rs is used with a threshold to adjust recall.

A large number of ML models are able to provide a probability estimate alongside predictions that can be used as the \rs. In \mytt{sklearn}, for instance, this is done with the \mytt{predict\_proba} method. Our implementation takes this approach. It is not necessarily the case that any given attack can generate a \rs. In these cases, every \rs value is the same, and recall must be 1.0.

\subsection{Selecting targets}
\label{sec:selecting-targets}

The \textit{target selection function} is used to distinguish between cases where all individuals are a priori understood to be equally attackable (\textit{mode=random}) versus where individuals with certain attributes, for instance outliers like individuals with a very high salary, are understood a priori to be more attackable (\textit{mode=pre-targeted}). In the former case, all individuals are potential targets, and in the latter the target selection function selects the attackable subset. The targets are then randomly shuffled and attacked in shuffled order.

\subsection{Example}

A data custodian plans to use a synthetic data tool to anonymize their data, but suspects that the tool does not protect outliers. One of the attributes is \textit{salary}, which is assumed to be known to the attacker. The custodian also assumes that the age and marital status of the target are known, i.e. $KA$=[\textit{salary, age, married}]. The $select_{atk}$() function will select individuals with the highest salaries, and the selection \textit{mode} is \textit{pre-selected}. The custodian is concerned that an attacker might want to learn either the \textit{zip} or the political \textit{party} of the target. Therefore the custodian will run two attacks, one with $UA=zip$ and another with $UA=party$. In both cases, the prediction criteria is \textit{attribute = value}. For the baseline predictor $pred_{base}$(), the custodian chooses two logistic regressors with \textit{salary}, \textit{age}, and \textit{married} as features, and \textit{zip} or \textit{party} as the predicted attributes.

\subsection{Computing the baseline and attack measures}

For each attack scenario, the following steps are taken to compute precision and recall for both the attack and the baseline:
\begin{enumerate}
    \item \label{step:loop} Do the following for each target $t$ in the shuffled list of targets until the halting criteria are met. This generates $L_{ack}$ and $L_{base}$; two lists of \tup tuples for the attack and baseline measures respectively:
    \begin{enumerate}
        \item \label{step:attack} \textbf{Run the attack measure:} Run the attack predictor $pred_{atk}$(\xda, $KA_t$, $UA_t$) and add the resulting \tup tuple to $L_{atk}$.
        \item \label{step:baseline} \textbf{Run the baseline measure:}
        \begin{enumerate}
            \item \label{step:remove} Remove the target $t$ from \xdo to create a new dataset \xdop.
            \item \label{step:predict} Run the baseline predictor $pred_{base}$(\xdop, $KA_t$, $UA_t$), and add the resulting \tup tuple to $L_{base}$.
        \end{enumerate}
    \end{enumerate}
    \item \label{step:make_pairs} Make a set of one or more precision/recall pairs for both the attack and baseline measures. To generate a precision/recall pair:
    \begin{enumerate}
        \item \label{step:threshold} Set a \rs threshold for $L_{*}$ (where ${*}$ is \textit{atk} or \textit{base}). The entries in $L_{*}$ with \rs at or above the threshold are taken as predictions, while the remainder are taken as abstentions. Set $T_{*}$ as the number of true predictions, $F_{*}$ as the number of false predictions, and $A_{*}$ as the number of abstains (see Section~\ref{sec:halting-criteria}).
        \item \label{step:compute_p_r} Compute $P_{*}$ and $R_{*}$ (attack precision and recall) from $T_{*}$, $F_{*}$, and $A_{*}$ as described in Section~\ref{sec:compute-precision-recall}.
    \end{enumerate}
    \item \label{step:compute_pcr} For each precision/recall pair computed in step \ref{step:make_pairs}, compute the \textit{Precision Recall Coefficient} (PRC) using Equation~\ref{eq:prc}. The PRC is a composite score (see Section~\ref{sec:prc}).
    \item \label{step:compute_alc} Set \xprcab to the highest attack PRC, and \xprcbb to the highest baseline PRC. Compute the \textit{Anonymity Loss Coefficient} (ALC) from \xprcab and \xprcbb according to equation \ref{eq:alc_rel} (see Section~\ref{sec:alc}).
\end{enumerate}

Note that the baseline prediction (step~\ref{step:baseline}) is independent from the type of attack or the anonymization method. The prediction is based only on \xdop, which is not anonymized. As such, the baseline prediction is really just a classic learning prediction problem, and any learning or analysis technique can be used. The goal here is simply to make the best prediction.

When removing a target from \xdo to make a baseline prediction (step~\ref{step:remove}), if the dataset is time-series or otherwise has multiple records per individual, then all records pertaining to the target must be removed. If the dataset is a social network with links between individuals, then all links for targets must be removed.  Although step~\ref{step:remove} prescribes that only the target individual is removed from \xdo for each prediction, this would be inefficient if for instance an ML fitting algorithm needs to be run for each target. This inefficiency can be avoided if an ML model that doesn't overfit is used to make predictions. The complete set of targets can be removed together, exactly like a holdout set in ML models. The model need be fitted only once, after which predictions are made on each individual target. Both Giomi et al.~\cite{giomi2022unified} and Steinke et al.~\cite{steinke2023privacy} take this approach.


Enough attack and baseline measures (step~\ref{step:loop}) should be run to ensure that the precision and recall measures are statistically significant. In our implementation, we compute the Wilson Score Interval after each pair of measures, and stop taking measures when the intervals for all precision/recall pairs are small enough (by default, we use an interval of 0.1 with 95\% confidence), and that further predictions will not yield substantially better PRC scores (Section~\ref{sec:halting-criteria}).

\subsection{Computing precision and recall}
\label{sec:compute-precision-recall}

We refer to the set of targets for which attack and baseline measures are made in step~\ref{step:loop} as \textit{attempts}. In step~\ref{step:threshold}, a threshold is applied to the \rs values. The attempts above the threshold are considered \textit{predictions}, while those below the threshold are considered \textit{abstentions}. A higher threshold leads to more abstentions and therefore a lower recall and (usually) higher precision. By selecting multiple different thresholds, we generate multiple precision/recall pairs.

In step~\ref{step:make_pairs}, precision is computed. In actuality, we compute two precision values, the \textit{measured} precision and a \textit{probabilistic} precision, for both attack and baseline.

The measured precision $P^{meas}_{*}$ is computed as
\begin{equation}
    P^{meas}_{*} = \frac{true \; predictions}{all \; predictions} = \frac{T_{*}}{T_{*} + F_{*}} \label{eq:prec} \\
\end{equation}
where $*$ is $atk$ or $base$, $T_{*}$ is the number of true predictions, and $F_{*}$ is the number of false predictions.

The probabilistic precision is the midpoint of the precision confidence interval. In our implementation, we use the Wilson Score Interval to compute precision upper and lower bounds from $P^{meas}_{*}$. The probabilistic precision $P^{prob}_{*}$ is computed as:

\begin{equation}
    P^{prob}_{*} = P^{lower}_{*} + (P^{upper}_{*} - P^{lower}_{*})/2 \label{eq:prec_prob}
\end{equation}

We use the probabilistic precision for the PRC and ALC scores. The primary reason for this is to reduce ALC sensitivity to small measurement errors when both baseline and attack precision are very high. This is further discussed in Section~\ref{sec:alc}.

There are several ways to compute recall, depending on whether the selection mode is random or pre-selected. With random selection, the attempts are taken as representative of the whole population, and so recall $R_{*}$ can be computed only from the attempts:
\begin{equation}
    R_{*} = \frac{predictions}{all \; attempts} = \frac{T_{*} + F_{*}}{T_{*} + F_{*} + A_{*}} 
    \label{eq:recall}
\end{equation}
where $A{*}$ is the number of abstentions.

Pre-selected targets, on the other hand, are not representative of the whole population. In this case, one may wish to compute recall relative to the whole population, or relative to only the pre-selected individuals. In the former case, recall may be computed as:
\begin{equation}
    R_{*} = \frac{(N - A_{*}) N'}{N D}
    \label{eq:recall_whole}
\end{equation}
where $N$ is the number of attempts, $D$ is the total number of records in the dataset, and $N'$ is the total number of pre-targeted individuals. In the latter case, recall may be computed as:
\begin{equation}
    R'_{*} = \frac{N - A_{*}}{N}
    \label{eq:recall_pre}
\end{equation}

\subsection{Selecting \rs thresholds}
\label{sec:diff-abstain}

There can be as many \rs thresholds as there are distinct \rs values. A large number of thresholds and corresponding PRC measures is not necessary, in part because the granularity in recall values is finer than it needs to be, and in part because high \rs thresholds with very few predictions produces too wide a confidence interval. In this case, we bin the \rs values, and set thresholds between the bins. How many bins to use depends on whether high-precision/low-recall PRC measures are likely to be produced with more predictions (halting criteria). This is discussed in detail in Section~\ref{sec:halting-criteria}.

Because baseline \rs values not comparable with attack \rs values, the recall values for the baseline PRC's in general does not match those of the attack PRC's. Even where they do match, the abstentions for the baseline may be for different target individuals than the abstentions for the attack. This is not a problem. The goal here is to establish a \textit{statistical} baseline, not to literally determine if each successfully attacked individual was or was not predicted by the baseline. This is because in a real attack (versus one simulated by the data custodian), the attacker does not have the ground truth. The attacker only knows statistically how likely a given prediction is to be correct.

\subsection{Independence between individuals}
\label{sec:independence}

Section~\ref{sec:introduction} states that a dataset cannot compromise the privacy of individuals not in the dataset. Strictly speaking, this only holds if dataset non-members are independent from dataset members. If member Mary and non-member Nate have identical data, then the baseline prediction could inadvertently exploit this fact to make a better prediction of Nate's attributes.

This problem is common to every measure that computes a baseline by removing target individuals, not just ours. Since other measures use the anonymized dataset for the baseline, the anonymization itself, if it is strong, will tend to hide the dependencies. Since our approach uses the original dataset for the baseline, we must always be careful to ensure that dependent members don't unduly effect the measure.

If dependent records are rare, then the effect on the baseline precision will likewise be small. If for instance 1\% of non-member records have a dependent member record, the computed baseline will differ by only 1\% or so. If dependent records are common, for instance in banking data with joint accounts, then the dependent records can have a significant effect on the baseline precision. In this case, either the dependent records must either be discovered and removed, or the baseline must be computed with a classifier that is not sensitive to overfitting.

\begin{figure}[tp]
\begin{center}
\includegraphics[width=1.0\columnwidth]{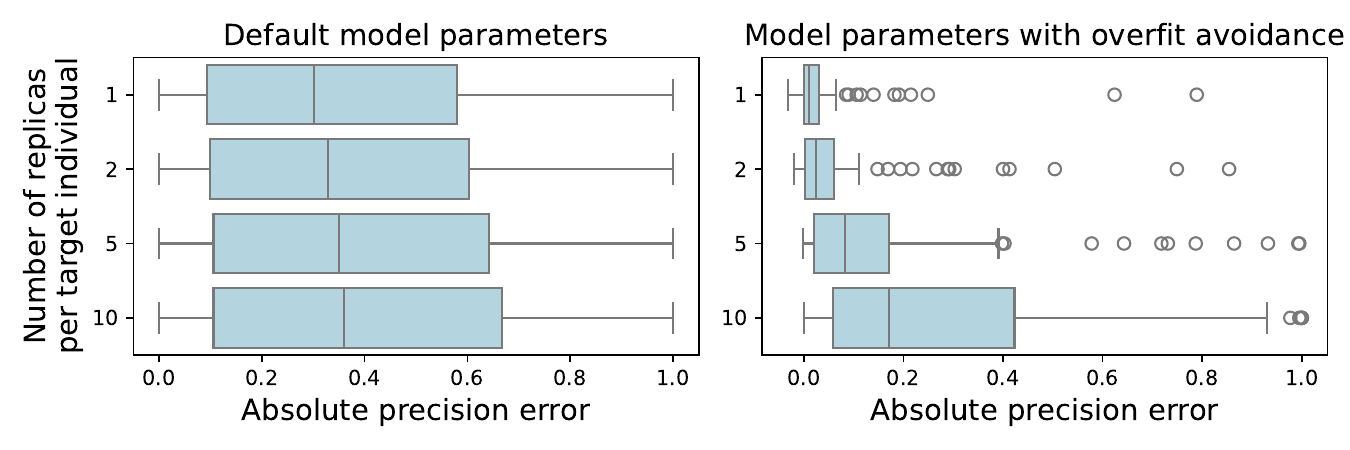}
\caption{Example of how dependent records can affect the baseline precision measure. The x axes show the absolute error between the baseline precision with the given amount of record replication and the baseline precision with no replication. The left plot shows the effect of dependent records when the default parameters of the RandomForestClassifier are used. The right plot shows the effect when the RandomForestClassifier is tuned to avoid overfitting.
}
\label{fig:ind_default_overfit2}
\end{center}
\end{figure}

Figure~\ref{fig:ind_default_overfit2} demonstrates this. We simulated the presence of replicated records and measured the difference in baseline precision between no replicated records (no dependence) and the case where non-member records are replicated in the member dataset. We experimented with 1, 2, 5, and 10 replicated records per individual. We made the measure using each of the categorical columns of the nine datasets in Table~\ref{tab:datasets} as the unknown attribute. There are 109 dataset/column measures in total, each contributing one data point to each boxplot. For each unknown attribute, we selected non-member records where the attribute value constituted less than 10\% of the dataset rows. We chose these rows because the precision of predictions on relatively infrequent values would in generally be correspondingly low, giving ample room for the precision of the replicated cases to be higher.

For each selected non-member record, we removed that record from the dataset. We then replicated the record 0, 1, 2, 5, and 10 times as members in the dataset. This simulates the case where \textit{every} non-member record is replicated, not just a few.  We used the \mytt{sklearn.ensemble} \mytt{RandomForestClassifier} to predict the unknown attribute. For each unknown attribute, we tested with 1000 non-members.

The left plot of Figure~\ref{fig:ind_default_overfit2} is for the case where the default parameters of the classifier are used. It demonstrates that dependent records can have a significant effect on the baseline precision measure. For the right plot, we modified some of the parameters to avoid overfitting, changing \mytt{n\_estimators} from 100 to 200, \mytt{min\_samples\_split} from 2 to 10, and \mytt{min\_samples\_leaf} from 1 to 10. This largely mitigates the effect of dependent records when there is one or two replicas per individual (which represents fairly extreme replication). However it comes at a cost. The absolute baseline precision of the right plot for 0 replicates is about 10\% lower than that of the left plot, which in turn means that the anonymity loss will be correspondingly overestimated.

We recommend that either dependent records be removed from the dataset before computing the baseline, or that the data custodian use a classifier that is not sensitive to overfitting.

\section{Composite measures}
\label{sec:composite}

The design in the previous section produces two sets of precision/recall measures. In order for data custodians to assess anonymity loss from the measures, they must be able to determine which among a set of precision/recall measures represents the most effective attack or the most accurate baseline. Does an attack with 90\% precision and 20\% recall represent more risk or less risk than an attack with 95\% precision and 5\% recall.

To answer this question, this section shows how to combine precision $P$ and recall $R$ into a composite score, conceptually similar to an F-score, and in particular how much weight to assign to $P$ and $R$.  We call this composite the \textit{precision-recall coefficient} (PRC).

The PRC allows us to do two things. First, it allows us to select the most effective attack and the strongest baseline. Second, it allows us to determine how much better or worse the attack is compared to the baseline, leading to a single measure, the \textit{Anonymity Loss Coefficient} (ALC), that quantifies the anonymity loss of a given attack on a given anonymized dataset.

\subsection{Precision-recall coefficient (PRC)}
\label{sec:prc}


A key goal in the design of PRC is to give data custodians adequate flexibility in how they weigh precision against recall. We let PRC be a value between 0 and 1, where 0 means no predictive power ($P=0$ or $R=0$), and one means perfect predictive power ($P=R=1$).

A central question is how much relative importance to assign to $P$ and $R$. Our intuition is that, unless $R$ is extremely low, $P$ should dominate. As a counter-example, imagine that we use F1 as our measure. This would mean that ($P=1$, $R=0.1$) and ($P=0.1$, $R=1$) are equivalent. This is clearly wrong: the former allows perfect predictions for 10\% of the dataset (a clear privacy risk), whereas the latter yields only 10\% correct predictions anywhere (almost certainly not a privacy risk).

As another example, consider the case of ($P=0.99$, $R=0.0001$): very good prediction for one in 10000 non-pre-targeted individuals (i.e. not known to an attacker in advance which individuals these might be). This attack would almost certainly be unattractive to an attacker because of the high probability of failure, or commensurately the high cost of success, and should have a low PRC.

%
%

These two examples suggest to us that precision is the more important of the two parameters, but that we cannot ignore recall entirely. We propose the following definition of PRC:


\begin{equation}
\label{eq:prc}
\text{PRC} =
\begin{cases} 
    \left(1 - \left(\frac{\log_{10}(R)}{\log_{10}(R_{\text{min}})}\right)^{\alpha}\right) \cdot P & \text{if } R > R_{\text{min}} \\
    R & \text{otherwise}
\end{cases}
\end{equation}

where \xcm and $\alpha$ are constants that allow a data custodian to control the effect of $R$ on PRC. The constant $\alpha$ determines how quickly the effect of $R$ decreases as it approaches \xcm. The constant \xcm is a threshold below which $P$ has no effect on PRC. Below this value, $PRC=R$. A conservative value is $R_{\text{min}}=0.0001$.

\begin{figure*}[tp]
\centering
\begin{subfigure}{0.30\textwidth}
    \includegraphics[width=\textwidth]{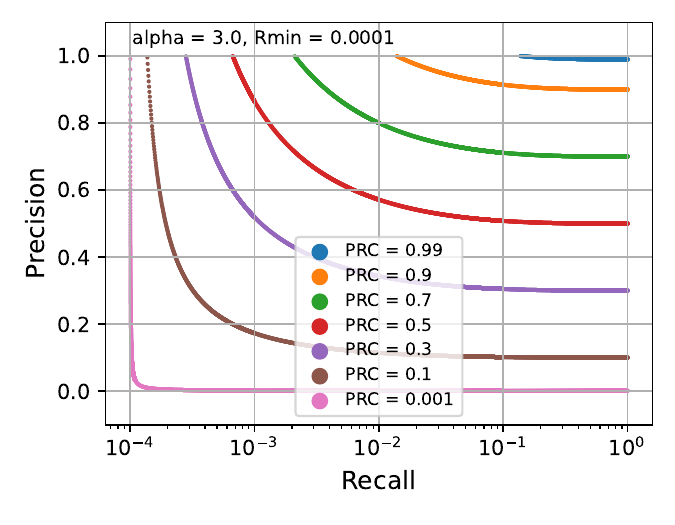}
    \caption{Curves of equivalent PRC for $\alpha=3$ and $R_{min}=0.0001$.}
    \label{fig:prec_recall_for_equal_prc}
\end{subfigure}
\hspace{0.02\textwidth} 
\begin{subfigure}{0.30\textwidth}
    \includegraphics[width=\textwidth]{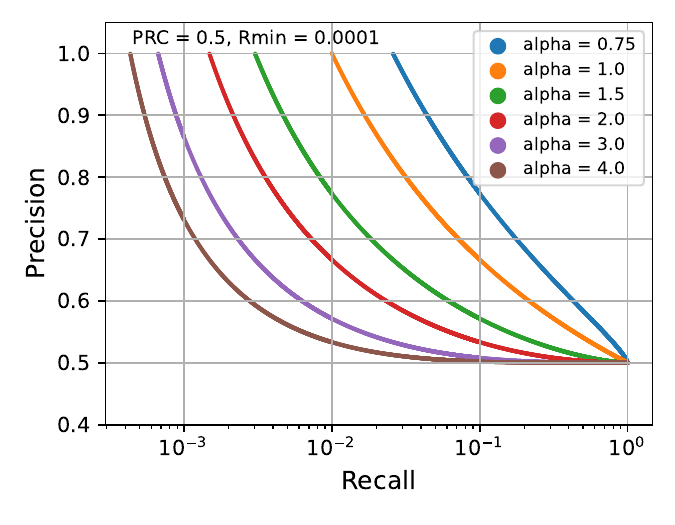}
    \caption{The effect of $\alpha$ on PRC. Higher $\alpha$ sharpens the knee of the curve.}
    \label{fig:prec_recall_for_diff_alpha}
\end{subfigure}
\hspace{0.02\textwidth} 
\begin{subfigure}{0.30\textwidth}
    \includegraphics[width=\textwidth]{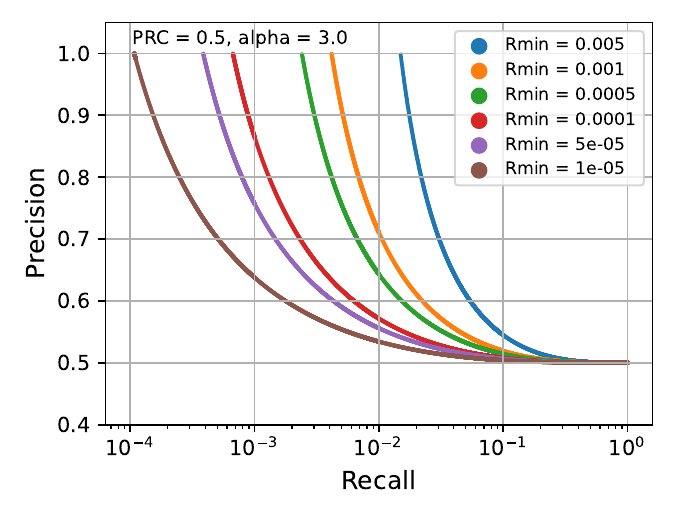}
    \caption{The effect of $R_{min}$ on PRC. Lower $R_{min}$ pulls the cross point to the left.}
    \label{fig:prec_recall_for_diff_rmin}
\end{subfigure}
\caption{PRC behavior}
\label{fig:combined_prc}
\end{figure*}

Figure~\ref{fig:combined_prc} illustrates the behavior of PRC. Figure~\ref{fig:prec_recall_for_equal_prc} shows curves of equivalent PRC for $\alpha=3$ and $R_{min}=0.0001$ from Equation~\ref{eq:prc}. Any point on any given curve has the same PRC.

Taking the curve for $PRC=0.5$ for example, the points at
\begin{itemize}
  \item ($P_{atk}=1.0$, $R_{atk} = 0.00148 = 1/676$)
  \item ($P_{atk}=0.6$, $R_{atk} = 0.02328 = 1/43$)
  \item ($P_{atk}=0.5$, $R_{atk}=1.0$)
\end{itemize}
would all be defined as having the same (relatively poor) attack effectiveness.

To be clear, we recognize that equating such extremely different values of $P$ and $R$ doesn't necessarily make intuitive sense in so far as they represent very different attack scenarios. Nevertheless, it is necessary to define some equivalence metric.

The constant $\alpha$ in Eq.~\ref{eq:prc} determines how sharp the knee of the equivalent-PRC curves are. Higher $\alpha$ leads to sharper knees, reducing the point at which $R$ has a significant effect on PRC.  This is illustrated in Figure~\ref{fig:prec_recall_for_diff_alpha}.

The constant $R_{min}$ in Eq.~\ref{eq:prc} determines where the PRC curve reaches the $P=1.0$ boundary. Lower $R_{min}$ pulls the equivalent-PRC curves to the left, reducing the point at which $R$ has a significant effect on PRC.  This is illustrated in Figure~\ref{fig:prec_recall_for_diff_rmin}.

Eq.~\ref{eq:prc}, with the parameters $\alpha$ and $R_{min}$, gives data custodians considerable flexibility in how to trade off the importance of precision and recall. Note that an alternative composite measure could have been the $F_\beta$-score, which is a weighted harmonic mean of precision and recall and has the advantage of being a well-known measure. The behavior of the $F_\beta$-score with respect to the problem of measuring anonymity is given in Appendix~\ref{sec:fbeta}. Our feeling is that the $F_\beta$-score does not have adequate tunability, especially given how little experience the community has with low-recall anonymity measures.

\subsection{Anonymity Loss Coefficient (ALC)}
\label{sec:alc}

The ALC is a value that quantifies how much better the attack performs over the baseline.  ALC is derived from the \xprca and \xprcb measures. After step~\ref{step:compute_pcr}, there are multiple PRC values for both the baseline and attack. In computing ALC, we use the maximum \xprca and \xprcb.

We define ALC so that $ALC>0$ if the attack is more effective than the baseline (there is some loss of anonymity), and $ALC<0$ if the baseline is more effective than the attack (there is no loss of anonymity). We define the maximum ALC as 1.0, which occurs when the attack has perfect precision and recall ($PRC_{atk}=1.0$), and the baseline is less than perfect.

It is important to note that any $ALC<1$ represents some attacker uncertainty, or equivalently some individual deniability. In other words, at any $ALC<1$, there is some amount of anonymity, and therefore there must be some threshold $ALC>0$ at which the data is considered acceptably anonymous.

$ALC$ can be defined as the difference between \xprca and \xprcb in either absolute or relative terms:

\begin{align}
    ALC_{abs} &= PRC_{atk} - PRC_{base} \label{eq:alc_abs}\\
    ALC_{rel} &= \frac{PRC_{atk} - PRC_{base}}{1-PRC_{base}} \label{eq:alc_rel}
\end{align}

An absolute measure is used by Yeom et al.~\cite{yeom2018privacy} and Stadler et al.~\cite{stadler2020synthetic} (called \textit{advantage}), while a relative measure is used by Giomi et al.~\cite{giomi2022unified} (called \textit{privacy risk}).

Both approaches have problems, and can produce substantially different measures. The table below gives three examples of PRC values and the resulting absolute and relative ALC values. The first two examples produce the same $ALC_{abs}=0.2$ even though the second example strikes us as being substantially worse privacy-wise. $ALC_{rel}$ captures this difference better.

\begin{center}
\begin{small}
\begin{tabular}{rll|ll}
    & \xprcb & \xprca & $ALC_{abs}$ & $ALC_{rel}$ \\ \hline
    1. & 0.1 & 0.3 & 0.2 & 0.22  \\
    2. & 0.75 & 0.95 & 0.2 & 0.8  \\
    3. & 0.99 & 0.999 & 0.009 & 0.9   \\
\end{tabular}
\end{small}
\end{center}

The third example shows that $ALC_{rel}$ is very sensitive to small differences when \xprca and \xprcb are both very high. It is not clear that the difference between a baseline of $PRC_{base}=0.99$ and an attack of $PRC_{atk}=0.999$ represents any meaningful privacy loss, especially given that the difference could easily be due to measurement error. Fortunately, this problem is mitigated in practice by basing PRC on the midpoint of the confidence interval rather than on the true measured precision. This prevents extremely high PRC values. For instance, if the attack loop is halted when the confidence interval reaches 0.1, then the best PRC value we can achieve is 0.95 (midpoint between 0.9 and 1.0).

Because of this mitigating factor, we prefer the use of the relative measure. In the remainder of this paper, we take $ALC$ to be $ALC_{rel}$ (Equation~\ref{eq:alc_rel}).

\begin{figure}[t]
\begin{center}
\includegraphics[width=0.85\columnwidth]{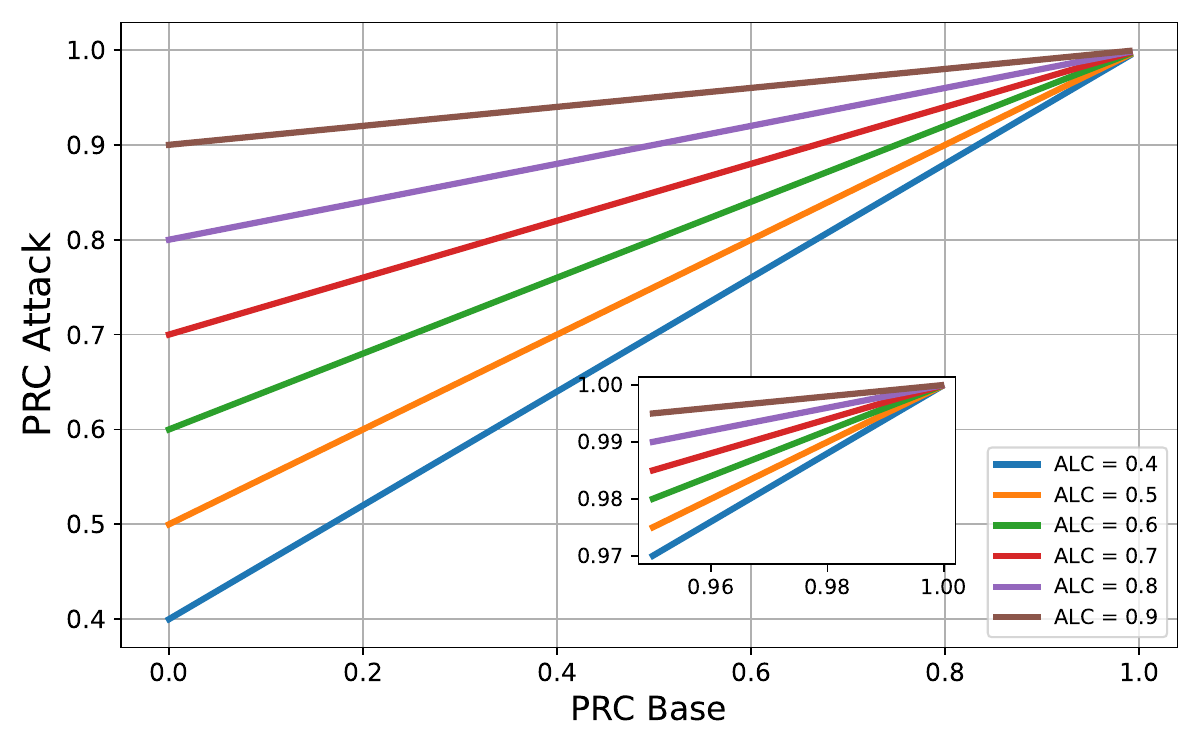}
\caption{Curves of equivalent ALC for different values of \xprca and \xprcb from Equation~\ref{eq:alc_rel}. From these curves, one may determine which ALC is safe for a given data release. Noting that PRC is equivalent to precision for a recall of 1.0, we believe that $ALC=0.5$ is a conservative threshold for strong anonymity.
}
\label{fig:alc_basic}
\end{center}
\end{figure}

Figure~\ref{fig:alc_basic} illustrates how ALC relates to different values of \xprcb and \xprca. A data custodian can examine these curves and determine what value of ALC is safe for a given data release (or tweak the PRC constants to produce ALC curves that work for the data custodian). We believe, however, that $ALC=0.5$ is a very conservative threshold. On this curve, when \xprcb is very low, say 0.05, \xprca is 0.525, barely better than a random coin toss. When \xprcb is fairly high, say 0.5, \xprca is 0.75, which represents substantial doubt for the attacker, or equivalently substantial deniability for the target.

When \xprcb is quite high, say 0.95, \xprca is 0.975 at $ALC=0.5$. It is difficult to argue in the abstract whether this represents a meaningful privacy loss or not, given that the baseline is in any event already quite high. Arguably if privacy-neutral predictions of 95\% confidence are possible, then the attribute is not very private even without the data release.
\ifblind
\else
Please see Appendix~\ref{sec:alc-threshold} for more discussion.
\fi

Nevertheless, a data custodian can always select a different ALC threshold for high values of \xprcb.

\begin{figure}[t]
\begin{center}
\includegraphics[width=0.85\columnwidth]{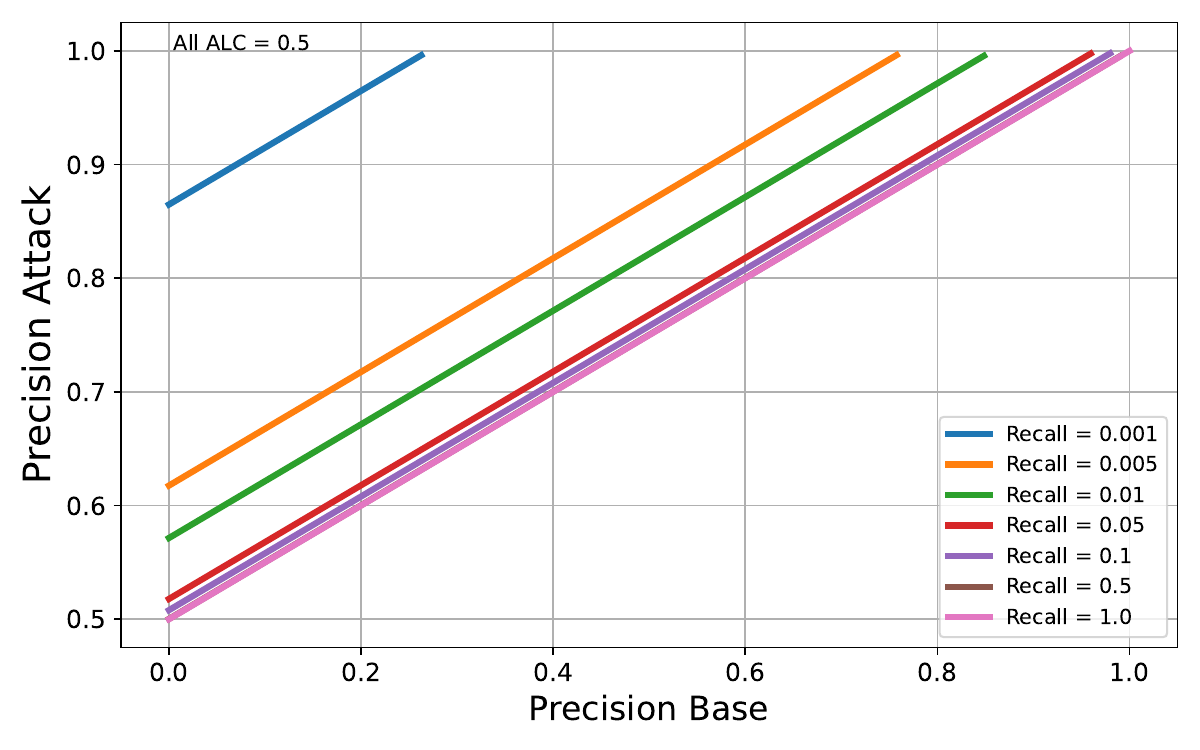}
\caption{Curves of equivalent ALC for different recall levels for $ALC=0.5$, $\alpha=3$ and $R_{min}=0.0001$. This shows that recall has very little effect until it is quite low, around 1/100.
}
\label{fig:alc_recall}
\end{center}
\end{figure}

Figure~\ref{fig:alc_recall} illustrates what levels of attack precision would be considered safe for different recall levels, given the conservative threshold of $ALC=0.5$. As expected, recall has little effect on ALC until it is quite low. At recall around 1/100, we see that the attack precision \xpa has to be only about 10\% higher to produce the same ALC as when recall is 1.0. Recall must be substantially lower before it has a very strong effect on ALC. At recall of 1/1000, we regard a precision improvement from $P_{base}=0.1$ to $P_{atk}=0.92$ as still being strongly anonymous. The rationale here is that an attacker would be extremely unlikely to bother making an attack with a 1/1000 success rate, and therefore even the individuals that happen to be vulnerable to the attack are effectively protected.

\ifblind
\else
\subsection{A (possibly large) set of ALC measures}
\label{sec:alc-set}

A single ALC measure is narrow in scope: it pertains to a single type of attack on a specific set of unknown attributes given a specific set of known attributes for a given dataset. The anonymity of an anonymized data release ultimately is defined by a set of ALC measures and their associated component parts (precision and recall scores). One might wish to condense the set of ALC measures into one or a small number of final measures. One obvious approach would be to take maximum ALC measure as the final measure. Alternatively, one could take the number of attacks whose ALC measure exceed a certain threshold. How best to deal with a set of ALC measures is out of scope for this paper, and a good topic for future work.
\fi

\subsection{Risk Assessment}
\label{risk-assessment}

The ultimate goal of an anonymity measure is to help assess risk and make decisions about data releases. We believe that precision and recall are good metrics for assessing risk because they are intuitive measures: what can one learn, with how much certainty, and for how many individuals? Of course, PRC and ALC per se are not very intuitive, but at least they derive from intuitive metrics.

We wish to make it clear what a threshold like $ALC=0.5$ does and does not represent. It does not mean that any $ALC>0.5$ is unsafe for release. Rather it means that any $ALC<0.5$ is safe for release; no more consideration is required by the data custodian.

We envision a risk assessment whereby known attacks are run on a data release, and those resulting in an ALC above some threshold are flagged for examination. Our current implementation in the \mytt{anonymity\_loss\_coefficient} package supports this by producing a rudimentary summary report. The data custodian can then examine precision and recall for each flagged case for such factors as data sensitivity, the likelihood of an attacker obtaining the necessary known attributes, the harm done to individuals, and the benefit to an attacker versus the cost of carrying out the attack. This is an important topic for future work.

\section{ALC and attacks implementation}
\label{sec:implementation}

Our pip-installable open-source python package \\ \mytt{anonymity\_loss\_coefficient} serves several purposes:

\begin{enumerate}
    \item It implements the ALC measure.
    \item It provides a framework for implementing new attacks.
    \item It contains a generic attack for microdata (synthetic data).
\end{enumerate}

The package supplies a basic attack loop as shown below.  From the caller's perspective, operation is quite simple. The caller creates the \mytt{ALCManager} object (line 1), enters the \mytt{predictor} loop to get rows to attack (line 2), runs the attack on each row (line 3), gives the \mytt{ALCManager} its prediction (lines 5 or 7), and then reads the attack results when the loop ends (line 8). The following provides more detail.

{\small
\begin{verbatim}
1  alcm = ALCManager(orig, anon)
2  for row in alcm.predictor(known, unknown):
3    pred,rank,abs = your_attack(row, alcm.anon)
4    if abs:
5      alcm.abstention()
6    else:
7      alcm.prediction(pred, rank)
8  df_results = alcm.results()
\end{verbatim}
}

The \mytt{ALCManager} class is initialized with the original dataset \mytt{orig} and a list of one or more anonymized datasets \mytt{anon} (line 1). It supports multiple anonymized datasets because often anonymized data is released as multiple datasets.  The \mytt{ALCManager} class identifies continuous columns and creates new columns that are discretized versions of the continuous columns. It then encodes categorical columns as integers in preparation for the random forest classifier. These pre-processed datasets are made available through the API (i.e. \mytt{alcm.anon}).

Despite the fact that our approach can support any True/False prediction logic (Section~\ref{sec:predictors}), the package itself currently supports only simple \mytt{pred=value} predictions. The purpose of the discretization is to accommodate \mytt{pred=value} predictions on continuous columns. Note also that the package currently does not support pre-targeted individual selection.

The caller starts the attack loop (line 2). The \mytt{predictor} is an iterable that returns rows to attack. The \mytt{predictor} first determines if a single value dominates the column. When this is the case, the rows with this value are down-sampled so that they appear in 50\% of the returned rows. This gives a better-balanced precision/recall measure at the expense of not accurately conveying a true random selection of rows.

The \mytt{predictor} method shuffles the original dataset, removes the first block of either 1000 rows or 10\% of rows, whichever is smaller, and trains a random forest classifier on the remaining rows. The \mytt{predictor} method then selects one row at a time. If the row needs to be down-sampled, then it continues to the next row, and in this fashion yields one \mytt{row} at a time from the block. The caller executes an attack on the yielded \mytt{row} using the preprocessed anonymized dataset \mytt{alcm.anon} (line 3). The attack returns either a predicted value and associated \rs, or an indication to abstain. If the former, then the prediction is conveyed in the \mytt{prediction} method (line 7). If the latter, then the abstention is conveyed in the \mytt{abstention} method (line 5).

For each yielded row, the \mytt{predictor} makes a baseline prediction on the \mytt{row} using the random forest classifier and associated prediction probability \mytt{proba} as the \rs. If the caller abstains, then the baseline prediction is used for the attack with a \rs of 0.0. After recording both the baseline and attack predictions and \textit{rank\_scores}, the \mytt{predictor} periodically checks the halting criteria to determine whether or not to exit the loop. If all of the rows in the block have been yielded without halting, then the \mytt{predictor} removes the next block from the original dataset, trains a new random forest classifier, and continues the loop. This continues until either the halting criteria are met or all rows are exhausted. Note that as an optimization, the halting criteria are checked only once in every (default) 20 loop iterations.

Once the loop halts, the \mytt{ALCManager} makes available two results dataframes to the caller. One dataframe contains every individual prediction. The other dataframe contains the precision/recall measures and their associated PRC and ALC scores (line 8), including the final ALC measure taken from \xprcab and \xprcbb, as well as other statistics.

\subsection{Halting criteria}
\label{sec:halting-criteria}

As more and more predictions are made, the number of predictions in the list $L_{atk}$ of attack tuples \tup grows. The larger $L_{atk}$ grows, the deeper the \rs threshold can be placed while still obtaining a small enough confidence interval. The basic idea behind the halting criteria is to continue making predictions so long as additional predictions are likely to generate low-recall measures with higher PRC scores than the higher-recall measures.

Our basic algorithm is:
\begin{itemize}
    \item Initialize the $N_{prc}$ counter to (default) 3.
    \item Make predictions. After each prediction:
    \begin{itemize}
        \item Set thresholds so as to generate $N_{prc}$ PRC measures.
        \item \textbf{If not} all PRC measures are significant, \textbf{continue}.
        \item \textbf{Else, if} the PRC score improves in each of the three lowest-recall measures by (default) 1\%, increment $N_{prc}$ and \textbf{continue}.
        \item \textbf{Else, end}.
    \end{itemize}
\end{itemize}

On top of this basic algorithm, there are several additional criteria. First, if there is only one or two distinct \rs values, then it is not possible to create three PRC measures. In this case, we halt once all one or two PRC measures are significant. In addition, there are two conditions under which early halting can occur, one where the ALC is clearly very low, and one where it is clearly very high. In the former case, so long as both attack and base confidence interval are less than 0.5, we compute ALC using the upper bound of the attack confidence interval and the lower bound of the base confidence interval. This give us an extremely optimistic ALC measure. If this measure is nevertheless less that (default) $ALC=0.4$, we halt on the grounds that anonymity is definitely safe. In the latter case, we use the upper bound of the base confidence interval and the lower bound of the attack confidence interval. This gives us an extremely pessimistic ALC measure. If this measure is greater than (default) $ALC=0.9$, we halt on the grounds that anonymity is definitely compromised.

\subsection{Best Row Match attack}
\label{sec:brm-attack}

The attack currently provided in the package is the \textit{best row match} attack. This attack is a general-purpose attack that can be used on any anonymized microdata or synthetic dataset, including those released as multiple datasets. An advantage of this attack is that it is a simple attack that any recipient of anonymized data might think to try: find the row in the data that best matches a known person, and infer unknown attributes from that row. A disadvantage is that, as a general-purpose attack, it does not target specific vulnerabilities that any given anonymization technique might have.

The attack runs by computing the Gower distance~\cite{gower1971general} between the known attributes of the target individual and those of each row in every table of the anonymized dataset that contains the unknown attribute and at least one known attribute.  The Gower distance is a value between 0 and 1, where 0 means that the two rows are identical and 1 means that they are completely different.  If the anonymized dataset does not contain all of the known attributes, then the missing attributes are treated as a mismatch and each assigned a value of 1.  All rows across all datasets that share the minimum Gower distance are considered to be matches. The prediction is the modal value of the unknown attribute among the matching rows

A \rs is computed for the prediction. The \rs takes into account two factors:
\begin{enumerate}
    \item The minimum Gower distance $G_{min}$ itself. The lower the distance, the better the \rs.
    \item The fraction of rows among the matching rows that have the modal value of the unknown attribute. The smaller the fraction of rows with the modal value, the less confident we are that we have the correct prediction, and the lower the \rs.
\end{enumerate}

Given these factors, we compute the \rs for a given prediction as:

\begin{equation}
    rank\_score = (1-G_{min}) * (M/C)
\label{eq:gower_score}
\end{equation}

where $M$ is the number of matching rows with the modal value of the unknown attribute, and $C$ is the total number of matching rows.

We have not yet made an effort to optimize this \rs equation; this is the first attempt. Section~\ref{sec:prior-approaches} shows that this approach certainly has some effectiveness. As future work, we can explore improvements on the \rs for this attack.

\subsubsection{Comparison with other best row match attacks}
\label{sec:other-brm-attacks}

Two prior best row match attacks are Giomi et al.~\cite{giomi2022unified} and Stadler et al.~\cite{stadler2020synthetic}. Similar to us, Giomi et al. uses the Gower distance. By default it bases its prediction on a single row even if multiple rows match. It does not take the quality of the match into consideration.  Stadler et al. use the best match only if it is a perfect match. If it is not, then similar to us they fall back on an ML prediction, with the difference that the ML model is trained over the anonymized data rather than the original data. Unless anonymization is extremely weak (i.e. many matches in the anonymized data are perfect), Stadler et al.'s approach does not do significantly better than the baseline.
\section{Comparison with prior approaches}
\label{sec:prior-approaches}

In Section~\ref{sec:related}, we describe a number of prior approaches that use a non-member baseline~\cite{yeom2018privacy, kassem2019differential, stadler2020synthetic, giomi2022unified, abowd2023census}. Using the notation from the description of our approach in Section~\ref{sec:design}, we can describe all of these techniques as having essentially the following approach to computing the baseline:

\begin{enumerate}
    \item[] \label{step:ploop} \textbf{Compute the baseline:} For each target:
    \begin{enumerate}
        \item \label{step:premove} Remove the target from \xdo to create a new dataset \xdop.
        \item \label{step:panon} Anonymize \xdop to create \xdap.
        \item \label{step:ppredict} Run the \textit{attack} predictor $pred_{atk}$(\xdap, *).
    \end{enumerate}
\end{enumerate}

There are four important differences between this baseline and the corresponding Step~\ref{step:baseline} from our approach in Section~\ref{sec:design}:
\begin{enumerate}
    \item The prior work does not take recall into account.
    \item The prior work makes the baseline prediction on an anonymized non-member dataset \xdap rather than the original dataset \xdop.
    \item The prior work reuses the attack predictor $pred_{atk}()$ rather than an independent best-performance predictor $pred_{base}()$.
    \item The prior work has an extra anonymization step.
\end{enumerate}

The second and third differences lead to a baseline that is certainly no better than ours, and may often lead to a poorer baseline. Making the prediction on an anonymized dataset \textit{necessarily} leads to a baseline that is no better, on average, since the anonymized dataset hides or distorts information. Using the attack itself as the baseline predictor certainly can do no better than a ``best-performance'' predictor, because if it did then the attack predictor would be used as the best-performance predictor.

The extra anonymization step means that the prior approach is computationally more expensive than our approach. Indeed, while the corrected measure from the US Census Bureau states that a version of the prior approach would produce a better measure, they do not implement it on the grounds of its being too expensive~\cite{abowd2023census}.

\begin{table}[t]
\begin{center}
\begin{small}
\begin{tabular}{rrllll}
\toprule
Dataset & Rows & \multicolumn{3}{c}{Columns} & Source \\
  &  & Tot & Cat & Con & \\
\midrule
    bank churners & 10127 & 22 & 10 & 12 & Kaggle  \\ 
    child & 20000 & 20 & 20 & 0 & SDV  \\ 
    insurance & 20000 & 27 & 27 & 0 & SDV  \\ 
    alarm & 20000 & 37 & 37 & 0 & SDV \\ 
    national2019 & 27253 & 24 & 20 & 4 & SDNIST \\ 
    adult & 30000 & 15 & 10 & 5 & SDV \\ 
    news & 30000 & 59 & 40 & 19 & SDV \\ 
    credit & 30000 & 30 & 9 & 21 & SDV \\ 
    census & 30000 & 41 & 33 & 8 & SDV \\ 
\bottomrule
\end{tabular}
\end{small}
\end{center}
\caption{Datasets used in the comparisons with prior approaches. The national2019 dataset comes from the SDNIST synthetic data project run by the National Institute of Standards and Technology (NIST). The bank churners dataset comes from Kaggle. The others are from the Synthetic Data Vault (SDV).
These datasets may all be found at \datasetsurl.}
\label{tab:datasets}
\end{table}

\begin{figure*}[ht]
    \centering
    \begin{subfigure}[b]{0.32\textwidth}
        \centering
        \includegraphics[width=\textwidth]{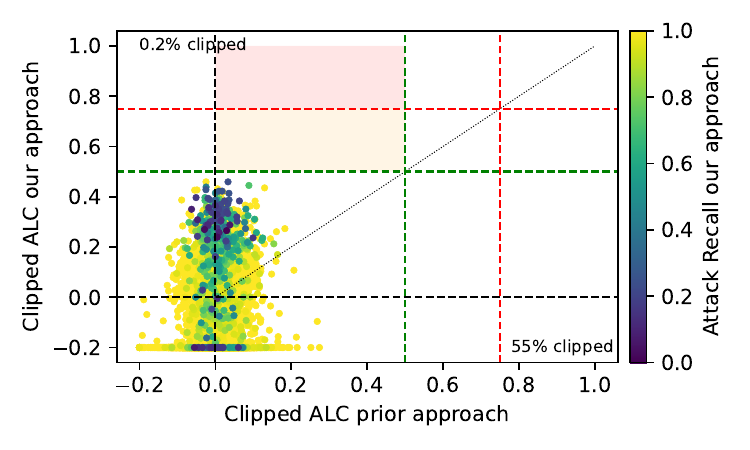}
        \captionsetup{width=0.9\textwidth} 
        \caption{Ours vs. prior approach, strong anonymization (80\% swapped). Both approaches regard the data as strongly anonymous.}
        \label{fig:alc_ours_vs_prior_strong}
    \end{subfigure}
    \begin{subfigure}[b]{0.32\textwidth}
        \centering
        \includegraphics[width=\textwidth]{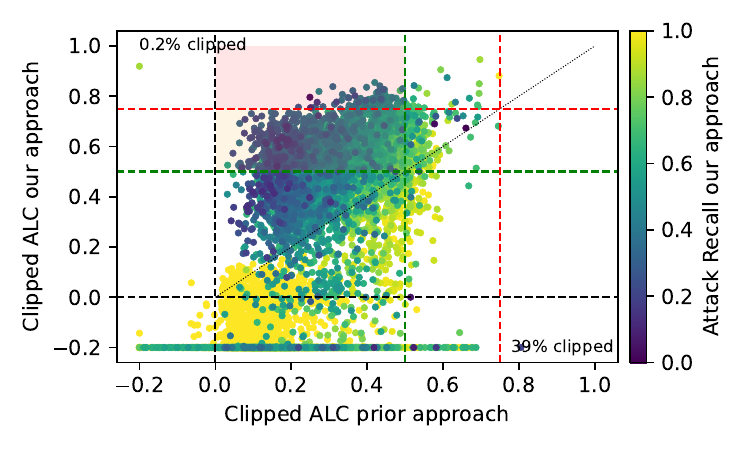}
        \captionsetup{width=0.9\textwidth} 
        \caption{Ours vs. prior approach, weak anonymization (20\% swapped). Over 25\% of prior approach attacks are false negatives.}
        \label{fig:alc_ours_vs_prior_weak}
    \end{subfigure}
    \begin{subfigure}[b]{0.32\textwidth}
        \centering
        \includegraphics[width=\textwidth]{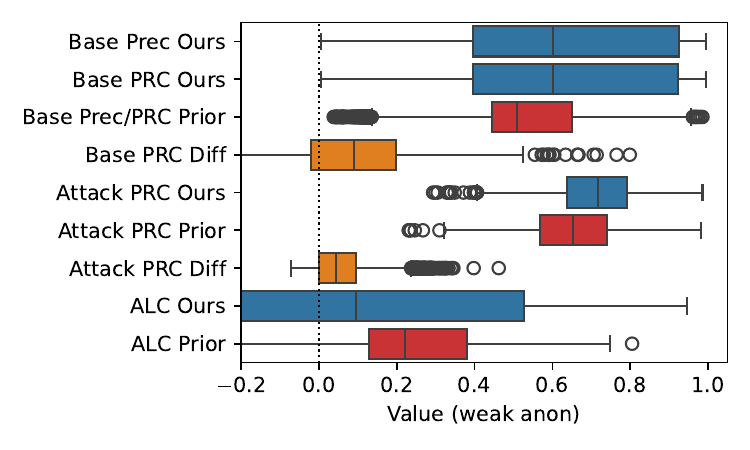}
        \captionsetup{width=0.9\textwidth} 
        \caption{A variety of base and attack measures, and their differences (ours - prior), from the attacks on the weak dataset from Figure~\ref{fig:alc_ours_vs_prior_weak}.}
        \label{fig:boxplots_weak}
    \end{subfigure}

    \caption{Comparison of our approach with prior approaches. The colorbar represents the attack recall values for attacks with recall. The prior approach both overestimates (below the diagonal) and underestimates anonymity.}
    \label{fig:overall_comparison}
\end{figure*}

\subsection{Our approach versus prior approach}

Figure~\ref{fig:overall_comparison} compares our approach with a version of the prior approach that is similar to Giomi et al.~\cite{giomi2022unified}. The purpose of this evaluation is not to evaluate the strength of attacks on realistic anonymization methods per se, but rather to compare our approach with the prior approach on an apples-to-apples basis. We ran measures on the 9 datasets shown in Table~\ref{tab:datasets}, which lists the number of rows, the number of continuous and categorical columns, and where the datasets can be found. These represent a wide variety of table characteristics.  The datasets were limited to 30K random rows for computational efficiency due to the brute-force approach taken by the best row match attack. From each dataset we generated two anonymized microdata datasets, one with strong anonymization and one with moderate anonymization (labeled ``weak''). Following~\cite{shlomo2010data} and \cite{spicer2009balancing}, we use random swapping of values within columns to generate anonymized datasets for comparison purposes. For strong anonymization, 80\% of values were swapped. For weak, 20\% were swapped.

We used each column in each dataset in turn as the unknown attribute. We selected 5 sets of other columns as known attributes. The known attributes were selected so as to minimize the number of known attributes while still ensuring that most of the value combinations were unique. Within these constraints, we selected random known attributes. In other words, we selected unknown attributes which were more likely to be unique and therefore attackable. Recall that when unknown attributes are continuous, they are discretized it into 20 bins and treated it as categorical. (When used as known attributes, continuous columns are not discretized.)

For each attack configuration (dataset + known attributes + unknown attribute), we ran an attack with our approach and one with the prior approach. In both cases, we used the best row match attack from Section~\ref{sec:brm-attack}. Our approach is described in Sections~\ref{sec:design} through \ref{sec:implementation}. For the prior approach, we used the following modifications:
\begin{itemize}
    \item The baseline was computed using the best row match attack rather than an ML model.
    \item The baseline was run against the anonymized dataset with targets removed instead of the original dataset.
    \item Recall is always 1.0. Because of this, the ALC measure for the prior approach is very similar to the Privacy Risk measure of Giomi et al.~\cite{giomi2022unified}.
\end{itemize}

In total, we ran 8883 attacks (known/unknown attribute combinations) with each anonymized dataset averaging 8 unknown attributes per attack. For the strong anonymization, we averaged 412 predictions and 1.2 precision/recall measures per attack. For weak anonymization, we averaged 491 predictions and 3.0 precision/recall measures per attack. The increased number of predictions and precision/recall measures for weak anonymization reflects the fact that more low-recall measures were explored for weak anonymization.

Each point in Figure~\ref{fig:alc_ours_vs_prior_strong} is for one attack on strong anonymization. ALC scores in the figure are clipped at -0.2 to make the figure more readable. Since ALC values below 0 in any event represent no anonymity loss, the details of data below 0.0 are not particularly interesting. The plots show what fraction of the data was clipped.

ALC values below the black dashed line ($ALC=0$) have no anonymity loss. ALC values below the green dashed line ($ALC=0.5$) have anonymity loss but we conservatively consider them safe. Values above the red dashed line ($ALC=0.75$) are considered to represent serious anonymity loss. Values between the green and red dashed lines are considered potentially at risk. The yellow shaded area represents cases where the prior approach regards the attack as safe, while our approach regards the attack as potentially at risk. The red shaded area represents cases where the prior approach regards the attack as safe, while our approach regards the attack as having serious anonymity loss. These are the cases where the prior approach would incorrectly conclude that the attack is safe when it may well not be.

The main takeaway from Figure~\ref{fig:alc_ours_vs_prior_strong} is that both our approach and the prior approach show that the data as strongly anonymous in that no attacks have an ALC greater than 0.5. In this regard, the two approaches draw the same conclusion for every attack.

\begin{table}[t]
\begin{center}
\begin{small}
\begin{tabular}{cc|cc}
\toprule
 Ours & Prior & No recall & Complete \\
\midrule
At risk & Safe & 16.54\% (1470) & 24.24\% (2154) \\
Serious & Safe & 0.59\% (52) & 1.24\% (110) \\
Safe & At risk & 0.6\% (53) & 1.52\% (135) \\
Safe & Serious & 0.06\% (5) & 0.01\% (1) \\
\bottomrule
\end{tabular}
\caption{Percentage (number) of attacks where the prior approach classifies anonymity incorrectly.}
\label{tab:wrong_conclusion}
\end{small}
\end{center}
\end{table}

Figure~\ref{fig:alc_ours_vs_prior_weak}, which is for weak anonymization, tells a different story. As Table~\ref{tab:wrong_conclusion} shows, over 25\% of the prior approach's attack classifications are incorrect. Most of these are false negatives: over 24\% of the attacks are attacks which our approach regards as at risk (yellow shading, 24.2\%) or serious (red shading, 1.2\%), but which the prior approach regards as safe. For these attacks, the mean error in the prior approach is 0.25 with standard deviation of 0.12. The 90th percentile error is 0.44. 1.5\% of the prior approach's attacks are false positives, where the prior approach regards the attack as at risk even though our approach shows them as safe. Note that the false negatives come from 6 of the 9 datasets: they are not the result of a single dataset being particularly problematic.

As can be seen in Figure~\ref{fig:alc_ours_vs_prior_weak}, the attacks for which our approach has substantially higher ALC scores also have lower recall. This is because the attack confidence score correlates with attack effectiveness, allowing the attack to find higher-precision, lower-recall attacks.

Both Figures~\ref{fig:alc_ours_vs_prior_strong} and~\ref{fig:alc_ours_vs_prior_weak} show that, compared to our approach, the prior approach sometimes has lower ALC scores (points above the diagonal), and sometimes higher. A higher prior ALC should not be interpreted as the prior approach simply being more cautious but otherwise valid.  Regardless of whether the prior ALC is lower or higher, it is a less accurate measure of anonymity. When the prior ALC is lower, relative to our approach, it is underestimating anonymity, and when it is higher, it is overestimating anonymity.

This can be seen from Figure~\ref{fig:boxplots_weak}. The base PRC for our approach is generally higher than for the prior approach, which means it is a more accurate measure. Better ML modeling would improve the base PRC for our approach more. Likewise, the attack PRC for our approach is also generally higher and therefore more accurate (modulo the 0.1 confidence bounds). Improvements in the attack increase the ALC score, while improvements in the baseline decrease the ALC score. Whether the prior ALC score is higher or lower depends on the relative magnitude of improvements in the attack and baseline. Either way, our approach represents a more accurate measure of anonymity loss.

\subsection{Effect of recall versus no recall}

We are interested in the question of how much of the improvement in our approach is due to the prior approach not using recall, and how much is due to the prior approach using attacks on the anonymized data as a baseline. To determine this, we ran the attack using our approach for the baseline (ML model over the original data), but only used the PRC score with recall=1.0. This is shows in Table~\ref{tab:wrong_conclusion} under the ``No recall'' column. 17.1\% of the attacks are regarded by the no-recall variant as safe while our approach with recall regards them as at risk (16.5\%) or serious (0.6\%). Likewise there are a few false positives (0.07\%). This shows that, while the lack of recall has a significant effect, it is not the only reason for the difference between our approach and the prior approach.
\section{Summary and future work}
\label{sec:summary}

This paper presents the design and implementation of anonymity loss measures for attribute inference attacks. It improves on prior work by incorporating recall into the measure, and by using a more accurate and efficient inference baseline which is derived from ML models over the original dataset minus the target, rather than running the attack on the anonymized dataset minus the target. We show that, for an example of moderately weak anonymization, over 25\% of the attacks that the prior approach found to be safe are actually at risk or serious risk when our approach is used. This paper also describes \mytt{anonymity\_loss\_coefficient}, an open-source, pip-installable Python package of our approach. The package implements the measure, has a framework for easily adding new attacks, and as of this writing contains one attack, the best row match attack on microdata.

As future work, we would like to improve the ML model used for the baseline, for instance by running several models and selecting the best one. It would also be interesting to look into ways of automating the discovery and removal of dependent records. We would like to apply our approach to real data releases, and in particular get feedback from data protection authorities on how to interpret ALC scores for the data releases as well as how to set PRC parameters.

There are a number of improvements that should be made to the package itself, such as adding more attacks, adding more types of predictions (multi-attribute inferences, ranges, etc.), and improving the best row match attack, for instance by exploring different \textit{rank\_scores}.  We invite researchers to contribute attacks to the package, and data custodians to use the package to evaluate their own data releases.

\textbf{Availability:} The code and data that produced the results in this paper, as well as the source code for the \mytt{anonymity\_loss\_coefficient} package, is available at \codeurl.



\newpage
\bibliographystyle{ACM-Reference-Format}
\bibliography{../../masterBib/master}

\clearpage
\appendix

\ifblind
\else
\section{Discussion}
\label{sec:discussion}

In this section, we discuss two related topics that shed additional light on our approach.

\subsection{Assumptions underpinning our approach}

It is instructive to examine how the assumptions between the prior approach and our approach differ with respect to computing the baseline PRC.

The prior baseline models a scenario where the attacker knowledge is limited to the known attributes of the targets and what can be learned from the anonymized data. No other general knowledge is assumed. Since the anonymized dataset is in fact released, this correctly models the minimum information the attacker may have. The baseline is therefore conservative in that it assumes an attacker with no external general knowledge.

By contrast, our baseline models a scenario where the attacker knowledge is the known attributes and what can be statistically learned from the original dataset. Since the attacker does not literally have access to the original dataset, our baseline represents the statistical information that the attacker \textit{could in principle} have rather than the minimum information that the attacker is definitely known to have. This baseline nevertheless is privacy neutral and so represents a safe baseline.

Either model can be inaccurate in terms of what new information a given attacker literally learns. Counter to the prior model's assumptions, the attacker may in fact have some external statistical information that improves the attack effectiveness in a privacy-neutral way. Likewise the attacker in our model may have less knowledge than we assume, or for that matter may have more knowledge than we assume because of some higher-quality external source of statistical information.

The advantage of our model is that it produces on average a higher baseline PRC, and therefore any given attack will be modeled as less effective (lower ALC). Given that our model is privacy neutral, a lower ALC can allow data custodians to justifiably reduce the amount of distortion in the anonymization or to release data that otherwise might have been withheld.

Our design in Section~\ref{sec:design} tacitly assumes that:
\begin{itemize}
\item That the attacker's known attributes are taken from the attributes of the released anonymous dataset.
\item That the baseline is computed from the known attributes of the original dataset from which the anonymized dataset is derived.
\end{itemize}

These assumptions are convenient because they allow the data custodian to compute baselines using the materials at hand: the original and anonymized data. In principal, however, neither of these design assumptions need be true. Attackers may well exploit other sources of information to improve their predictions. When it is clear that this is the case, if the data custodian can likewise exploit the other sources of information to improve the baseline, then they should do so.

\subsubsection{Example: the US Census Bureau's reconstruction attack}
\label{sec:uscensus}

A real example of this comes from the US Census Bureau's (USCB) reconstruction attack, where the attacker is trying to infer race and ethnicity. In measuring the effectiveness of the attack, the USCB assumed that the attacker has publicly available data about targets as the known attributes. This data included the target's name, address, age, and sex. Likewise, the data collected by the USCB includes these attributes.

As part of its anonymization process, the USCB removes the name altogether, and converts the address into a block number. Nevertheless, the original data available to the USCB includes name and address, and therefore could in principle be used as part of the baseline analysis. Likewise, the attacker has access to name and address. There is no reason for the attacker to throw away the name, as the USCB assumes. Although the attacker does derive the block from the address, there is no reason for the attacker to subsequently ignore address in its attack, again as the USCB assumes.

As it so happens, names can often be used to infer race and ethnicity. Surnames work particularly well for Asian and Hispanic individuals~\cite{elliott2009using}, while given names can often distinguish black and white races~\cite{fryer2003causes}.

Address can also be used to improve race/ethnicity inference. Consider, for instance, a block A that is 30\% black and 70\% white. If the block B to the west of this block is predominantly black, while the block C to the east is predominantly white, the attacker can infer that individuals with addresses on the west side of block A are more likely to be black.

Exploiting name and address will improve the quality of both the baseline and the attack. If one is improved more than the other, then this will produce a different anonymity loss measure than the one where name and address are ignored. Given that the USCB has chosen to disregard name and address, the vulnerability measures most recently reported by the USCB~\cite{abowd2023census} may still be overly pessimistic.

%
%
%

\subsection{Lack of an evidential basis for assigning risk thresholds}

In deciding whether to release anonymized data, data custodians must decide on some threshold. This is true of any anonymization measure: a threshold of K for K-anonymity, a threshold of $\epsilon$ or other parameters for differential privacy, and so on. In the case of our approach, a threshold for ALC must be set, and the PRC parameters $\alpha$ and $C_{min}$ must be chosen as well.

Setting anonymization measure thresholds is a general problem. There is no consensus for instance on how to choose a value of $\epsilon$ for differential privacy~\cite{dwork2019differential}. Likewise, El Emam cites a variety of data disclosure guidelines that specify the minimum cell size for aggregated data (i.e. K-anonymity), where cell size is the number of individuals that constitute a aggregate statistic (\cite{el2020evaluating} appendix 1 supplementary material). The cell size guidelines vary from 3 to 30.

One of the reasons why setting thresholds is difficult is that, to our knowledge, there are no known reports of \textit{malicious} attacks on anonymized or even pseudonymized data. By malicious, we mean genuine ``black hat'' attacks in the wild (as opposed to white hat attacks motivated by academics or journalists). There is, therefore, no evidential basis on which to establish anonymity thresholds. Data custodians must ultimately rely on predicting the costs and benefits to attackers.

%
%

\subsection{On the choice of $ALC=0.5$ as a threshold}
\label{sec:alc-threshold}

Despite the lack of an evidential basis for assigning thresholds, we suggest that $ALC=0.5$ is a conservative threshold given $\alpha=3$ and $C_{min}=0.0001$. This was briefly discussed in Section~\ref{sec:alc}. Here we expand on this choice of threshold. We argue that, at $ALC=0.5$, one of three conditions, or some combination of the three conditions, always holds:
\begin{enumerate}
\item The attacker has substantial uncertainty that a prediction is correct (or equivalently, the target has substantial plausible deniability).
\item The baseline precision is in any event high, and so any increased attack precision is of negligible additional value to the attacker (or equivalently, is of negligible additional harm to the target).
\item The probability of a high-precision attack is very low, and so the target is unlikely to be attacked, and indeed the attacker likely lacks incentive to run the attack in the first place.
\end{enumerate}

\begin{table}[ht]
\centering
\begin{tabular}{lllll}
\toprule
 & Recall & $P_{\text{base}}$ & $P_{\text{atk}}$ & \\
\midrule
1 & 1.0 & 0.05 & 0.54 & Substantial deniability \\
2 & 1.0 & 0.9 & 0.95 & Diminishing returns \\
3 & 1.0 & 0.99 & 0.995 & Diminishing returns \\
4 & 0.001 & 0.2 & 0.96 & Low attack probability \\
\bottomrule
\end{tabular}
\caption{Example recall and precision values when $ALC=0.5$.}
\label{tab:alc-example}
\end{table}

Regarding the first condition, assuming recall is one, at $ALC=0.5$ and a low base precision of $P_{base}=0.05$, then $P_{atk}$ is only 0.54 (see the first item in Table~\ref{tab:alc-example}). The attacker's prediction is barely better than a coin toss. If we regard $p_{atk}=0.9$ as representing sufficient deniability for the target, then at $P_{base}=0.05$, even the resulting $ALC=0.89$ could still be regarded as anonymous.

Suppose on the other hand that we regard $P_{atk}=0.95$ as providing some but nevertheless insufficient deniability. At $ALC=0.5$, a base precision of $P_{base}=0.9$ is needed. In this case we suggest that the added value to the attacker of going from a precision of 0.9 to 0.95 is of limited value. Likewise for the target, the \textit{additional} loss of deniability has little impact. The same argument could be made going from $P_{base}=0.99$ to $P_{atk}=0.995$.

When recall is very low, then at $ALC=0.5$ it is possible to have high attack precision even with low baseline precision. For instance, at $R_{base}=R_{atk}=0.001$, if $P_{base}=0.2$, then $P_{atk}=0.96$. Here we argue that the low probability of being the target (1/1000) represents adequate protection for targets. Indeed, the incentive for the attacker to run the attack at all is low, given the low probability of success. Given this, all targets are safe simply because no targets are attacked.

Of course, it is up to data custodians and data protection authorities to determine if $ALC=0.5$ is a suitable threshold and whether $\alpha=3$ and $C_{min}=0.0001$ are suitable parameters for their data release, taking into account the sensitivity of the data, the difficulty of obtaining the information required to run the attack, the motivation of attackers, and so on. 
\section{GDPR Analysis}
\label{sec:legal}

We are interested in the question of whether the European Union data protection regulation GDPR supports the idea of a baseline inference.
\ifblind
\footnote{Temporary note: this section is written by a legal scholar who specializes in GDPR data anonymity}
\fi
At its core, GDPR Article 4(1) defines personal data as:
\begin{noindent}
\begin{quoting}
\textit{any information relating to an identified or identifiable natural person (`data subject')}
\end{quoting}
\end{noindent}
Identifying includes by direct means (e.g. name, account number) and indirect means (e.g. zip code). While GDPR only applies to personal data (Article 2 (1), recital 26) defines anonymous as:
\begin{noindent}
\begin{quoting}
\textit{personal data rendered anonymous in such a manner that the data subject is not or no longer identifiable is not considered personal data.
}
\end{quoting}
\end{noindent}

The first question is, does GDPR support the core concept that if an individual is not present in a dataset, then the release of that dataset does not violate that individual's privacy (i.e. is not considered personal data with respect to that individual)? Since the very definition of personal data requires the presence of an `identified or identifiable natural person' in the data, we take it as self-evident that the lack of presence means that the person cannot be identified and therefore the data cannot violate the person's privacy.

The second question is, does GDPR support the concept of a baseline inference? The GDPR itself intentionally says very little about how to determine if data is anonymous, and in particular says nothing about the role of inference in anonymity. The Article 29 Data Protection Working Party, however, in its ``Opinion 05/2014 on Anonymisation Techniques'', does address inference~\cite{article29}. It identifies inference as one of three criteria against which anonymization can be evaluated:

\begin{noindent}
\begin{quoting}
     \textit{Inference, which is the possibility to deduce, with significant probability, the value of an attribute from the values of a set of other attributes.}
\end{quoting}
\end{noindent}

A key phrase here is ``significant probability''. The Article 29 opinion elucidates what it means by ``significant probability'' through several examples. Primary among these is the example of k-anonymity and the Homogeneity Attack~\cite{machanavajjhala2007diversity}. This is an attack where, for a given set of known attribute values, there is only one possible value for the secret attribute. In this case, the probability of a correct inference is 100\%, which is indeed significant.

Nevertheless, the criteria does not say ``100\% probability'', so it implicitly recognizes a probability less than 100\% to potentially break anonymity. In discussing the Permutation mechanism (swapping values between records), the Article 29 opinion says this:

\begin{noindent}
\begin{quoting}
\textit{not knowing which attributes have been permutated, the attacker has to consider that his inference is based on a wrong hypothesis and therefore only probabilistic inference remains possible.}
\end{quoting}
\end{noindent}

In other words, the Article 29 opinion recognizes a ``probabilistic inference'' which does not break anonymity. Though the Article 29 opinion does not elucidate how to determine when an inference is merely probabilistic, versus when it breaks anonymity, this example suggests that GDPR supports the general idea of a baseline inference.

It is important to recognize that the baseline inference is a probabilistic measure: the precision as measured across a \textit{group of individuals}. The behavior relative to any single given individual is, however, not probabilistic.

Suppose there is an individual target $T$ where the inference prediction made by an analysis on the baseline dataset for $T$ (as a non-member) is incorrect, but an inference prediction for $T$ made by attack on the anonymous dataset (where $T$ is a member) generates a correct prediction. Seemingly this particular individual's privacy has been compromised, even if the statistical attack precision is below the baseline precision.

Here we rely on the uncertainty of the attacker. While the prediction for $T$ went from wrong to right in point of fact, the attacker doesn't know this. The uncertainty of the attacker protects the target. This leads to the third question: does the Article 29 opinion recognize uncertainty as a legitimate form of protection?

The answer is `yes'. For instance, when discussing inserting noise, the Article 29 opinion says:
\begin{noindent}
\begin{quoting}
\textit{even if the noisy disclosure mechanism is known in advance, the privacy of the data subject is preserved, since a degree of uncertainty remains.}
\end{quoting}
\end{noindent}

The baseline essentially establishes a degree of uncertainty which by definition must be anonymous relative to a given individual, since that individual is not present in the dataset from which the baseline was derived. If the degree of uncertainty is the same or greater when the individuals are present in the anonymized dataset, then the individuals' privacy is equally or better protected. Since this is true for every individual in the anonymized dataset, the dataset may be regarded as anonymous by GDPR standards, and therefore non-personal data.

We believe that our approach can make a contribution to the definition of anonymity in the GDPR and other privacy regulations. By defining a baseline inference, it can be stated that any anonymization technique that only allows inferences below this baseline is certainly anonymous, at least with respect to the known inference attacks.


%
%
%
%
\section{Towards a universal measure of anonymity}
\label{sec:other-measures}

Attribute inference is only one of several measures that are commonly used in the literature to evaluate anonymity. In this section, we argue that many of these measures are redundant to attribute inference.

\subsection{Reidentification}

A common measure of anonymity for techniques like pseudonymization or sampling that do not distort the data per se is reidentification rate. This is the fraction of individuals that can be correctly identified in the ``anonymized'' dataset given certain known attributes. The harm of reidentification, however, is not the reidentification itself, but the fact that the unknown attributes of the identified individuals can be inferred. This is of course the attribute inference problem, and so reidentification rate itself is redundant to attribute inference.

\subsection{Similarity measures}

There are a class of privacy measures for synthetic data that measure the similarity between the synthetic data and the original data. A simple measure is the fraction of rows that match exactly. Others measure closeness, for instance the Hamming distance between synthetic and original data. A good overview of these techniques can be found in~\cite{ganev2023inadequacy}. The key problem with these metrics is that they don't take a baseline into account. For instance, a k-anonymous dataset with no suppression or generalization would yield perfect similarity with the original data even though the k-anonymized dataset is anonymous. Measuring attribute inference against a baseline effectively reflects closeness without the inherent problem of a pure closeness measure.

\subsection{Membership inference}

A common attack in the literature is membership inference (see Hu et al. for a survey~\cite{hu2022membership}). Ostensibly, the goal of membership inference is to determine if a given individual is in an anonymized dataset (usually the training data for an ML model). The implication is that knowledge of membership is privacy violating~\cite{shokri2017membership}. The attacks executed in this literature, however, determine if a given individual was randomly selected from a population of known individuals. Since simply being part of a random selection in and of itself is not privacy violating, the membership inference attack does not measure privacy loss in any meaningful way.

To make the point clear, suppose that the attacker is trying to determine membership in a training dataset for a model that contains only individuals with lung cancer. In this case, the population from which the members are sampled by definition all have lung cancer, and determining that an individual was randomly selected for the training dataset does not infer any new information.

\subsection{Singling out and linkability}

The European Article 28 Data Protection Working Party Opinion on Anonymization Techniques \cite{article29} gives three criteria for assessing anonymization, Singling out, Linkability, and (Attribute) Inference. Giomi et al. describes attacks for all three criteria. So far as we can tell, the motivation for doing so in Giomi et al. is not because each of the three attacks has distinct advantages, for instance each revealing information not revealed by the others, but rather simply to provide a complete set of attacks relative to the EU criteria. We find singling out and linkability to be largely redundant to attribute inference.

Singling out is ``\textit{the possibility to isolate some or all records which identify an individual in the dataset}.'' In the context of pseudonymized data, this is clearly a problem because the attacker can infer additional attributes about an individual from the isolated record. As implemented by Giomi et al., the attacker predicts that there is exactly one individual in the original dataset with a given set of attributes. This in an of itself is not obviously a problem because the attacker doesn't have access to the original data. It would only be a problem if the attacker could also infer additional attributes. Since this is in fact the attribute inference problem, the singling out attack is unnecessary.

Linkability is ``\textit{the ability to link, at least, two records concerning the same data subject or a group of data subjects (either in the same database or in two different databases)}.'' For the linkability attack, Giomi et al. posit an original dataset that is synthesized. The original dataset is then split into two column-disjoint datasets, A and B. The attacker has access to the synthetic dataset and A and B. Given knowledge of attributes that may be in both A and B (but do not uniquely identify rows in either A or B), the goal of the attacker is to link the rows in A and B that come from the same row in the original dataset. This is a pretty contrived setup, and in fact the attack mechanism is similar to the attribute inference attack, but separately applied to both A and B. The attacker finds the best matching rows in A and B, and then assumes the two are linked. The linking step is incidental. This could easily be recast as an attribute inference attack by subsequently predicting the unknown attributes by reading them from the selected rows in A and B without the linking step. We therefore find the linkability attack to be redundant to the attribute inference attack.
\ifblind
\section*{Appendix A: Using the F-beta function}
\addcontentsline{toc}{section}{Appendix A: Using the F-beta function}
\else
\section{Using the F-beta function}
\fi
\label{sec:fbeta}

In section~\ref{sec:prc} we define the Precision-Recall Coefficient (PRC), a composite measure for precision and recall. The PRC has two parameters, $\alpha$ and \xcm, which are used to adjust the sharpness of the PRC curve and the point at which it intersects the $P=1.0$ line.

An alternative would have been to use the $F_\beta$ function, which has the advantage of being a well-known precision-recall measure.  The $F_\beta$ function is defined as:

\begin{equation}
F_\beta = (1 + \beta^2) \cdot \frac{\text{Precision} \cdot \text{Recall}}{(\beta^2 \cdot \text{Precision}) + \text{Recall}}
\label{eq:fbeta}
\end{equation}

To help justify the use of a bespoke composite measure, we plot the behavior of the $F_\beta$ function for different values of $\beta$ in Figure~\ref{fig:fbeta}. For low values of $\beta$, the $F_\beta$ function does have the desired effect of favoring precision over recall until recall is very low. However, the ability to independently adjust both the sharpness of the PRC curve and the $P=1$ intersection point is lost.

\begin{figure*}[t]
\centering
\includegraphics[width=0.8\textwidth]{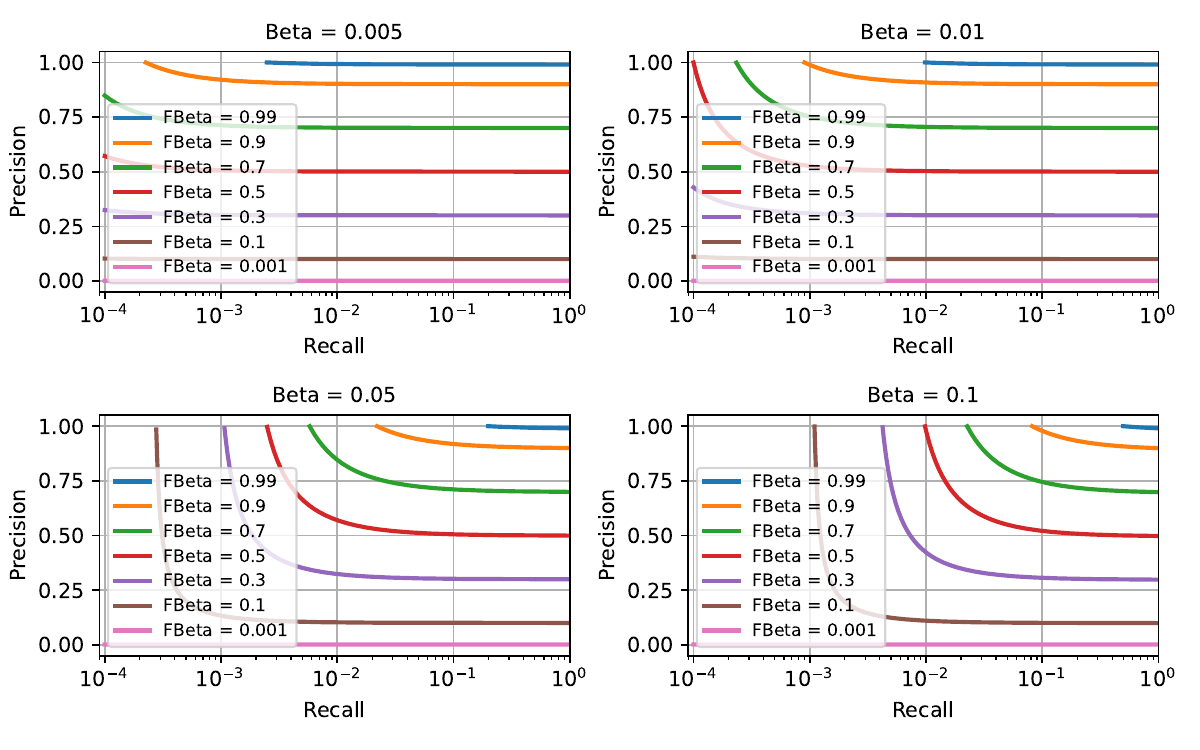} 
\caption{The effect of $\beta$ in the $F_\beta$ function.}
\label{fig:fbeta} 
\end{figure*}

\fi

\end{document}